\documentclass{aastex}
\usepackage{emulateapj5}

\usepackage{epsf}
\usepackage{onecolfloat}
\newcommand{\Om}{\Omega_m}
\newcommand{\Ob}{\Omega_b}
\newcommand{\LCDM}{\rm {\Lambda CDM}}

\newcommand{\DA}{D\!_A(z)}
\newcommand{\hz}{H(z)}

\newcommand{\Vsur}{V_{\rm survey}}

\newcommand{\Ox}{\Omega_X}
\newcommand{\hMpc}{h^{-1}{\rm\;Mpc}}

\newcommand{\trihGpc}{h^{-3}{\rm\;Gpc^3}}

\newcommand{\ihMpc}{h{\rm\;Mpc^{-1}}}
\newcommand{\kmax}{k_{\rm max}}
\newcommand{\kfit}{k_{\rm fit}}

\newcommand{\Vbox}{V_{\rm box}}
\newcommand{\Ro}{\rm Rho^2}
\newcommand{\fL}{f_{\rm L}}
\newcommand{\fNL}{f_{\rm NL}}

\begin{document}
\twocolumn[%
\submitted{Accepted by \textit{The Astrophysical Journal} 7-11-2005}
\title{Baryonic acoustic oscillations in simulated galaxy redshift surveys}
\author{Hee-Jong Seo, Daniel J. Eisenstein \\ \protect \today}
\affil{Steward Obsevatory, University of Arizona, 933 North Cherry Avenue, Tucson, AZ 85721}
\email{hseo@as.arizona.edu,deisenste@as.arizona.edu}

\begin{abstract}
Baryonic acoustic oscillations imprinted in the galaxy power spectrum
provide a promising tool for probing the cosmological distance scale and dark energy.  We present results from a
suite of cosmological N-body simulations aimed at investigating possible
systematic errors in the recovery of cosmological distances. We show the robustness of baryonic peaks against nonlinearity, redshift distortions, and mild biases within the linear and quasilinear region at various redshifts. While mildly biased tracers follow the matter power spectrum well, redshift distortions do partially obscure baryonic features in redshift space compared to real space. 
We calculate the statistical constraints on cosmological distortions from N-body results and compare these to the analytic results from a Fisher matrix formalism. We conclude that the angular diameter distance will be constrained as well as our previous Fisher matrix calculations while the Hubble parameter will be less constrained because of nonlinear redshift distortions.   

\end{abstract}


]

\section{Introduction}
Baryons create a distinct oscillatory signature in the power spectrum of the large-scale structure of the universe \citep{Peebles70,Bon84,Holtzman89,HS96,Eis98a}. These baryonic acoustic oscillations have been seen in the anisotropies of cosmic microwave background \citep{Mil99,deB00,Han00,HalDasi,Lee01,Netter02,BenoitArcheops,BennettWmap,Pearson03} and recently in large galaxy redshift surveys \citep{Eis05,Cole05}. Because the density of baryons is less than that of cold dark matter, the oscillations in the matter power spectrum are weaker in amplitude than those in the cosmic microwave background (hereafter CMB). However, whether in the CMB or in late-time structure, the oscillations define a constant comoving length scale in linear perturbation theory.

In our previous paper \citep[][hereater SE03]{Seo03}, we demonstrated that large galaxy redshift surveys can constrain the Hubble parameter and angular diameter distance to a precision of a few percent using the imprinted baryonic acoustic oscillations as a standard ruler. The physical scale of the oscillations can be determined from matter density and baryon density of the universe, which in turn are deduced from the shape and relative amplitude of the baryonic peaks in CMB anisotropy data \citep{Eht98,Eis03}. One can then compare the physical scale with the observed length scales of oscillations in transverse and line-of-sight directions in galaxy redshift surveys to yield the angular diameter distance and Hubble parameter at the given redshift. This in turn measures the evolution of dark energy as well as the spatial curvature and $\Om$. See \citet{Blake03}, \citet{Linder03}, \citet{Hu03}, \citet{Amen04}, \citet{Cooray04}, \citet{Dolney04}, and \citet{Mat04} for similar studies. 

Because of its weak amplitude modulations in matter distribution, the baryonic features are susceptible to the erasure by the nonlinear coupling of Fourier modes \citep{Jain94,Meiksin99,MW99,Sco99}, such as can result from nonlinear growth, nonlinear bias \citep{Kaiser87,Coles93,Sch98,Dekel99,Col99,Sel00,White01} or nonlinear redshift distortions \citep{Hamilton98,Hatton98,Sel01,White01,Sco04}. Nonlinear mode-coupling also produces additional small-scale power, thereby altering the shape of the power spectrum and potentially disguising the locations of baryonic features. It is therefore crucial to examine whether the oscillatory features are robust against various nonlinearities and whether the remaining ripples can effectively distinguish a dilation in distance from other non-cosmological effects in the clustering of galaxies. 

In this paper, we present an N-body study of the effect of nonlinear growth, nonlinear redshift distortions, and halo bias on the detectability of acoustic oscillations. Previously, \citet{Meiksin99} used N-body simulations to show the effects of nonlinearity on baryonic signatures in the present-day large-scale structure. They studied the effect of bias and redshift distortions for various cosmological models. Our study extends their work to higher redshifts. We adopt a $\LCDM$ model consistent with Wilkinson Microwave Anisotropy Probe (WMAP) data \citep{Spergel03} and generate density fields at redshifts of 3, 1 and 0.3. We then investigate and quantify the erasure of baryonic features in the matter power spectrum and biased galaxy power spectrum in real space and redshift space at those redshifts. We also attempt to remove the nonlinear alteration from the shape of the power spectrum and thereby recover the underlying contrast of baryonic features. As this work was being completed, related studies by \citet{Springel05} and \citet{Angulo05} appeared on this topic.

In  SE03, we used the Fisher information matrix to calculate predictions for the statistical constraints on dark energy. We used a conservative choice of nonlinear scale $\kmax$ (=$\pi/2R$) by requiring $\sigma_R \sim 0.5$ and excluded power in smaller scales from our analysis. For larger scales, we adopted a linear growth function, linear redshift distortions, and a linear bias model with an additive offset. In reality, as the transition from linear to nonlinear scales is not discrete, the effects of nonlinearity, i.e., mode coupling and more complicated scale-dependence, may mildly contaminate the power spectrum even on large scales. Here, we use our N-body simulations to assess the impacts of these nonlinear effects on the baryonic features on large scales relative to the statistical constraints we calculated in SE03. We compare our N-body results with the choices of nonlinear scale in SE03. Our results will provide further guidance for linear approximations in various studies with baryonic physics.

Finally we want to investigate the distance constraints available from galaxy surveys taking account of the full N-body effect. The nonlinear effects not only inhibit us from detecting the weak baryonic signatures on small scales but also increase the statistical variance of power spectrum above the Gaussian estimates. We perform $\chi^2$ analysis on our simulated power spectra to fit to the cosmological distances and compare the constraints with those in  SE03. The result will show whether we can deduce the information on cosmological distances from the power spectrum altered by the nonlinear effect.

In \S~\ref{sec:nbody} we describe the parameters of our cosmological N-body simulations. In \S~\ref{sec:matterps} we present the effect of nonlinear growth and redshift distortions on baryonic features in the matter power spectrum. In \S~\ref{subsec:nonlinminus} we remove the effect of nonlinearity on the broadband shape and study the resulting features. In \S~\ref{sec:bias} we study the baryonic signatures in biased power spectra in real space and redshift space. In \S~\ref{sec:chisq} we present the errors on cosmological distances resulting from $\chi^2$ analysis.

\section{Cosmological N-body simulations}\label{sec:nbody}

We run a series of cosmological N-body simulations using the Hydra code 
\citep{couch95} in collisionless ${\rm P^3M}$ mode.  
We use the CMBfast \citep{SeZa96,Za98,ZaSe00} linear power spectrum to generate many initial Gaussian random density fields at redshift of 49 and evolve them to redshifts of 3, 1, and 0.3. The cosmological parameters we use to generate the initial fields are $\Om=0.27$, $\Ox=0.73$, $\Ob=0.046$, $h=0.72$, and $n=0.99$. The initial fields are normalized by requiring $\sigma_8 = 0.9$ at $z=0$ and assuming a linear growth function. 
Each simulation box represents $\Vbox=512^3h^{-3}{\rm\;Mpc^3}$ and contains $256^3$ dark matter particles ($\sim 8.28 \times 10^{11} M_{\rm sun}$/particle). We compute gravity using $256^3$ force grids with a Plummer softening length of 0.2 $\hMpc$. 
We use 51 simulations at $z=1$ and 0.3, and 30 simulations at $z=3$. Note that the total volume of the simulations is much larger than the survey volume parameters listed in  SE03 so that we can study the effect of nonlinearity with little interference from statistical variance.

 The resulting density field of each simulation box is Fourier transformed, and the squared complex norms of Fourier coefficients are spherically averaged over all simulations to give the matter power spectra in the wavenumber-shells of widths $\Delta k = 0.005 \ihMpc$. A mode is included in a shell if its discrete wavenumber falls in the shell. Because each mode in the discrete transform is included in one and only one wavenumber-shell, the shells are not correlated for a Gaussian field even though the $\Delta k$ is smaller than the size of an independent cell in Fourier space, $2\pi /\Vbox^{1/3}$. Using small $\Delta k$ ensures that the contrast of narrow features are not artificially reduced. However, because a thin shell contains fewer modes, we do want to apply some smoothing. We use Savitzky-Golay filtering \citep{Press92}, as this can preserve peak heights better than boxcar smoothing. The Savitzky-Golay method also gives estimates of the derivatives\footnote{We use $4^{\rm th}$ order polynomial with a filter width of $\Delta k=0.04 \ihMpc$ to smooth power spectra and $\Delta k=0.05 \ihMpc$ to derive the derivatives of power spectra.}. From comparisons between power spectra before and after smoothing, we believe that this procedure neither introduce misleading baryonic features nor erase meaningful features. For power spectra in redshift space, we spherically average the Fourier transform of the density fields after displacing particles according to their peculiar velocities assuming a distant observer.

\begin{figure*}[t]
\epsscale{2}
\plottwo{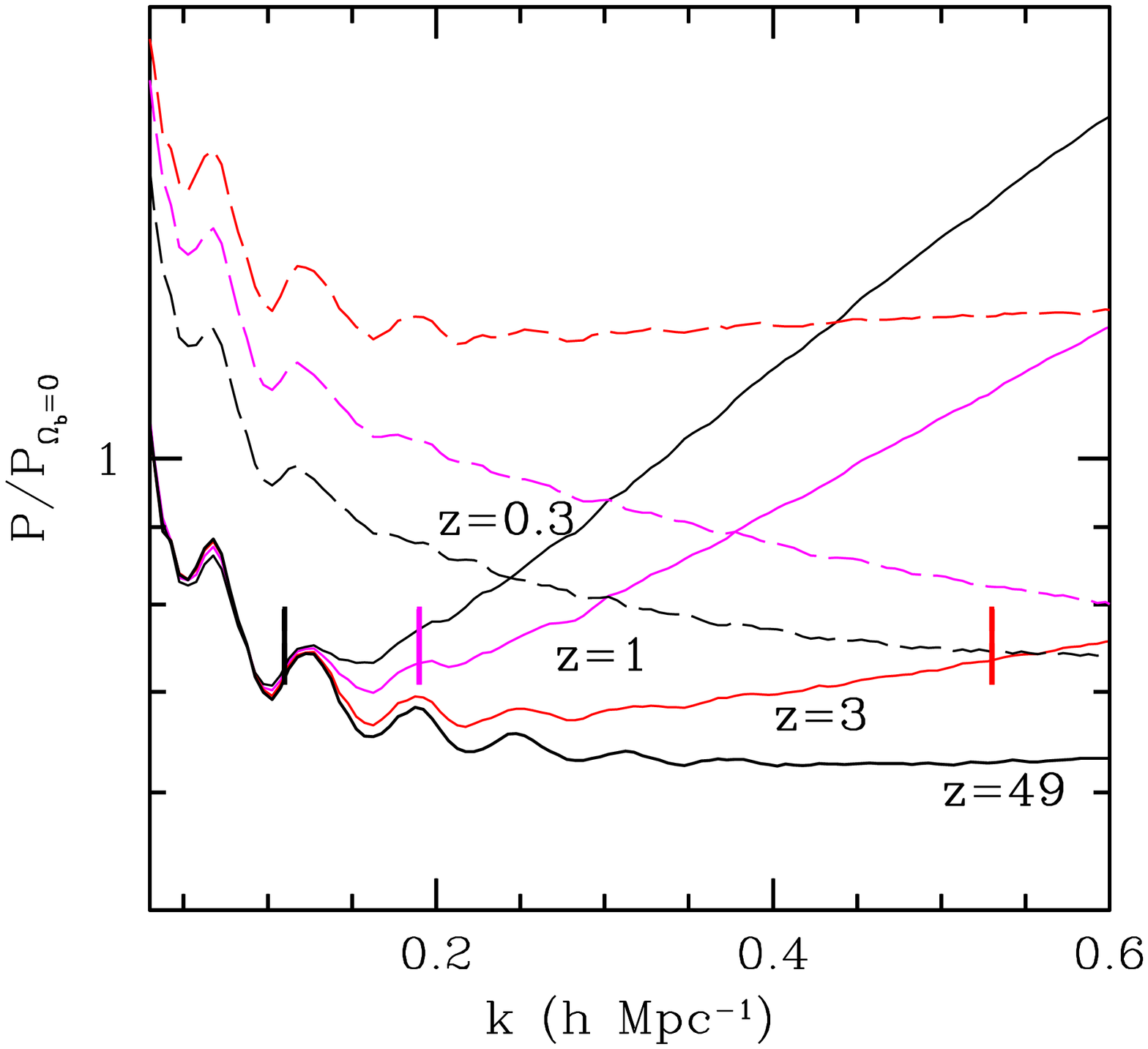}{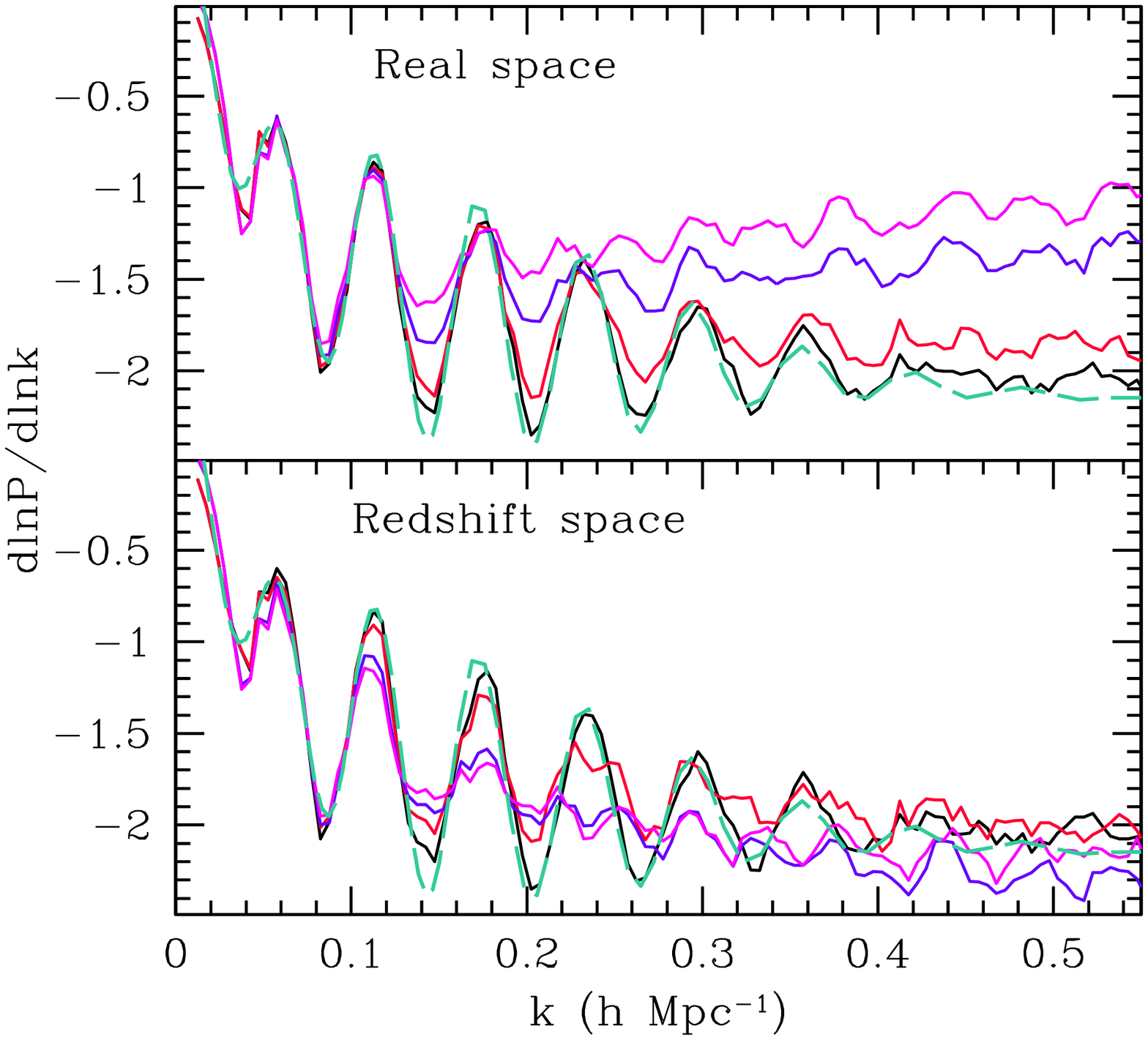}
%\plotone{temp.eps}
\caption{The matter power spectra at various redshifts. Left: each power spectrum is divided by a zero-baryon power spectrum at the given redshift. Solid lines are for the real-space clustering, and dashed lines with the same color are for the corresponding redshift-space clustering. Right: $d\ln P/d\ln k$ from the matter power spectra at various redshifts in real space (upper panel) and in redshift space (lower panel). Green dashed line: the input power spectrum from the CMBfast, black line: the linear matter power spectrum at $z=49$ generated from the input power spectrum, red: the nonlinear matter power spectra at $z=3$, blue: $z=1$ and violet: $z=0.3$. The first troughs in $d\ln P/d\ln k$ from the N-body results are lower than that of the input power spectrum due to the interaction of the Savitzky-Golay smoothing with the boundary at $k\sim 0$.}
\label{fig:dlnPlin}
\epsscale{1}
\end{figure*}

\begin{figure*}[t]
\epsscale{2}
\plottwo{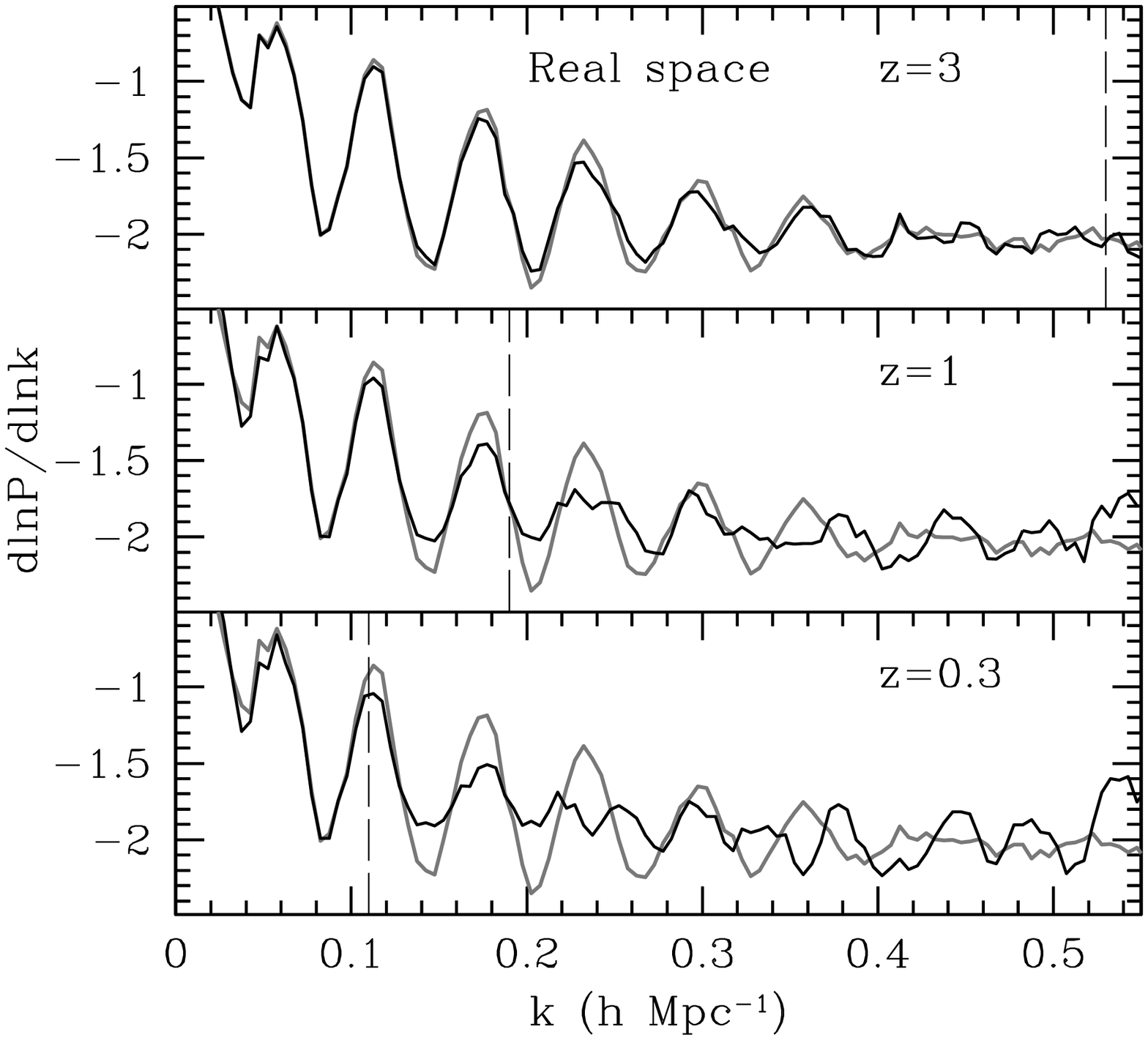}{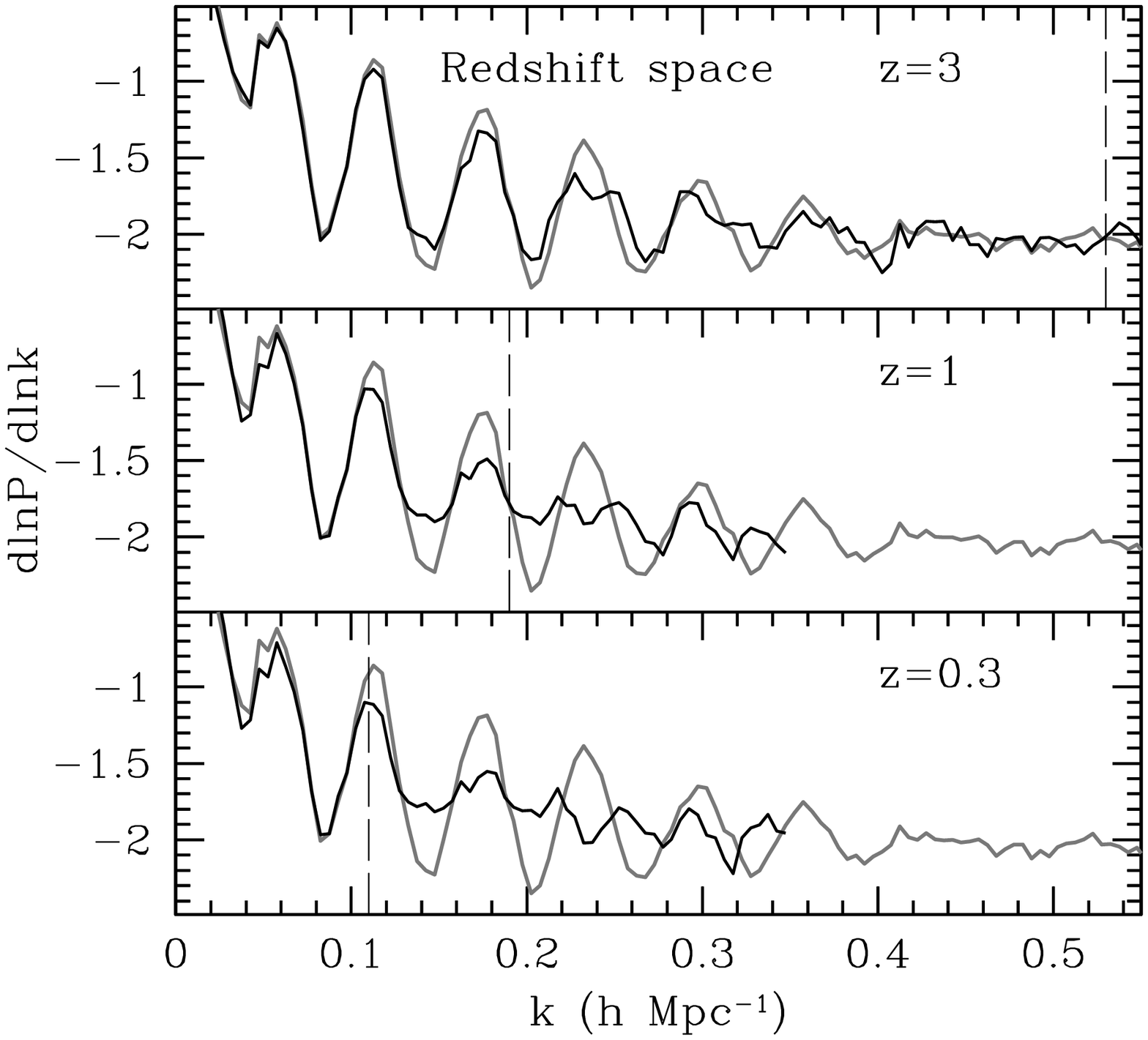}
\caption{Left: $d\ln P/d\ln k$ from $P_{\rm nonlinear}-\fL (c_0,k,k^2)$ at various redshifts in real space. Gray solid line: the linear matter power spectrum at $z=49$, black solid lines: the nonlinear matter power spectra at various redshifts. Right: the redshift-space $d\ln P/d\ln k$ from $P_{\rm nonlinear}-\fL (c_0,k,k^2)$ after corrected for the finger-of-God suppression. Note that the gray lines in this panel denote the linear matter power spectrum in {\it real} space at $z=49$ while black lines are for the nonlinear matter power spectra in {\it redshift} space. The vertical dashed lines denote the value of $\kmax$ we assumed in SE03.  We conclude that the baryonic features survive on large scales despite the nonlinear growth with the redshift distortions imposing an additional but mild degradation.}
\label{fig:dlnPlinminus}
\epsscale{1}
\end{figure*}

\section{The nonlinear matter power spectrum}\label{sec:matterps}
\subsection{Nonlinear effects in the matter power spectrum}\label{subsec:nonlin}
As mass perturbations on a given scale approach order unity in amplitude, linear perturbation theory breaks down and the gravitational growth of perturbations in one mode is increasingly coupled with perturbations in other modes. 
The higher-order contribution resulting from this mode coupling hinders the detection of features in the initial power spectrum, including the baryonic acoustic oscillations. That is, the additional power contributed from other modes blurs the initial features at a given Fourier mode as they mix with a convolution of other modes \citep[][and references therein]{Jain94}. Nonlinear growth from mode coupling increases power above the linear growth rate on small scales, resulting in a bigger statistical variance for any underlying initial features. As the amplitude of density perturbations grows with time, these nonlinear effects become stronger and proceed to larger scales.

As a basic model, one might distinguish the linear regime from nonlinear regime with a scale $R$ ($= \pi /2\kmax$) for an appropriate rms overdensity fluctuation, $\sigma_R$, at a given epoch, and assume scales with a smaller rms overdensity as linear. In studies of the statistical expectations from large redshift surveys, imposing more conservative values of $R$ improves the linear approximation, but this is at the expense of more of the remnant linear information in the quasilinear regime beyond $R$. On the other hand, more forgiving criteria for $R$ will result in an overestimation of the performance: the precision of the acoustic scale measurement improves as $\kmax$ increases, finally saturating beyond about $0.25 \ihMpc$ because of Silk damping (see Figure 4 of SE03). Only an N-body study can say whether this nonlinear scale accurately accounts for the erasure of features in the initial power spectrum such as baryonic oscillations. This will give an additional handle in the transition from linear to nonlinear regime beyond that traced by the increased amplitude. We will use the notation $\kmax$ in this paper to characterize the nonlinear scale and to assess the linear information from the baryonic features surviving nonlinearity.

Figure \ref{fig:dlnPlin} shows the effect of nonlinear gravitational growth on baryonic acoustic oscillations in matter power spectra from N-body simulations.
The solid lines in the left panel of the figure show the spherically averaged real-space power spectra divided by a zero-baryon power spectrum\footnote{A power spectrum generated from the fitting formula in \citet{Eis98a} for our fiducial $\Om$ and $h$ but $\Ob=0$} \citep{Eis98a} at various redshifts. The growth rate calculated from rms overdensity fluctuations at $16 \hMpc$, $\sigma_{16 \hMpc}$, is consistent with the linear growth rate, while the values of $\sigma_{8 \hMpc}$ show nonlinear effects at lower redshift. The vertical lines denote the nonlinear scales, $\kmax$, adopted in  SE03 to satisfy $\sigma_R \sim 0.5$ ($\kmax = 0.11\ihMpc$ at $z=0.3$, $0.19\ihMpc$ at $z=1$ and $0.53\ihMpc$ at $z=3$). 

As expected, nonlinear structure formation increases small-scale power and obscures small-scale baryonic features. The effect proceeds to larger scales with time. The difference in slope between the linear power spectrum at $z=49$ and the power spectra at lower redshift shows that remnant nonlinearity exists even for $k< \kmax$. At lower redshifts, the contrast of baryonic peaks on large scales is decreased because of the nonlinear mode-coupling.

The right panel of Figure \ref{fig:dlnPlin} shows logarithmic derivatives of the matter power spectrum with respect to wavenumbers, generated from Savitzky-Golay filtering. This plot of derivatives is useful not only because it effectively manifests fine details of power spectra but also because $d\ln P/d\ln k$ is what enters into the Fisher information matrix and creates the standard ruler test with baryonic oscillations. Of course, the derivative $d\ln P/d\ln k$ increases the noise in the power spectra despite our use of smoothing.

The real-space $d\ln P/d\ln k$ in Figure \ref{fig:dlnPlin} (upper panel) demonstrates that the oscillatory features are well distinguished for $k< \kmax$ at $z=0.3$ and $z=1$ despite the slight nonlinearity. At $z=3$, it becomes difficult to distinguish baryonic features beyond $k \sim 0.4 \ihMpc$ because the decreased contrasts of small-scale peaks caused by Silk damping \citep{Silk68} make them susceptible to even small amount of nonlinearity or noise. Fortunately, baryonic features in $k > 0.3 \ihMpc$ have minor contributions to cosmological information (SE03).

\subsection{The effect of redshift distortions}\label{subsec:redis}
Since we measure redshifts of galaxies rather than their physical distances, the three-dimensional galaxy power spectrum is subject to redshift distortions, the angle-dependent distortion in power spectra caused by the peculiar velocity of galaxies \citep{Kaiser87,Hamilton98,Sco04}. On large scales, the bulk motions of large-scale structure toward overdense regions enhance power, and on small scales, the virial motions within and among halos create an apparent extension along the line of sight, known as the finger-of-God effect \citep{deLa86}. This suppresses power on small scales. 
In linear theory, the large-scale power enhancement by the redshift distortions follows a simple form \citep{Kaiser87}. This is true only for the asymptotic limit of large scale, and in general, the nonlinear effect in velocity fields deviates the redshift-space power from Kaiser formula even on fairly large scales \citep{Sco04}. 
We seek to estimate how much the nonlinear effect of large-scale redshift distortions affects the baryonic features.

The dashed lines in the left panel of Figure \ref{fig:dlnPlin} show the redshift-space power spectra divided by a zero-baryon power spectrum. The figure depicts the progress of redshift distortions with nonlinearity and their effect on the baryonic features in the matter power spectrum. The redshift-space power spectra on large scales have a higher amplitude than that of real space by the amount predicted by linear theory \citep{Kaiser87} for the asymptotic limit. Even on large scales, we observe a slight suppression in redshift-space power with respect to linear theory, in agreement with \citet{Sco04}. On small scales, the finger-of-God effect not only suppresses the overall power but also decreases the contrast in baryonic features. As expected, the finger-of-God effect increases with time, and the resulting suppression makes the position of the second peak (which is beyond the nonlinear scale for $z=0.3$) appear slightly shifted. For our cases of study, the finger-of-God effect generally produces $10-20\%$ of suppression in redshift-space power at $\kmax$ in all redshifts when compared to the prediction by linear theory. 

The lower right panel of Figure \ref{fig:dlnPlin} gives a more clear view of the redshift distortions decreasing the contrasts and introducing noise at $k< \kmax$. Nevertheless, the baryonic features are still preserved up to $k \sim \kmax$ for $z=0.3$ and $z=1$, and  $k \lesssim 0.3 \ihMpc$ for $z=3$. The result at $z=0.3$ seems especially encouraging in that the features are preserved even beyond $\kmax$. 

It is important to note that these curves are spherically averaged power spectra in redshift space while we aim to use anisotropic information of power spectra in real observations. In the three-dimensional power spectra in redshift space, wavevectors nearly perpendicular to the line-of-sight direction will preserve baryonic features as well as the real-space power in  Figure \ref{fig:dlnPlin}, and wavevectors nearly along the line-of-sight direction will appear more smeared than the redshift-space power in the figure.

\subsection{Acoustic features after restoration of the broadband shape}\label{subsec:nonlinminus}
In this section, we consider the effect of nonlinear modification of the slope of the power spectrum on the baryonic features. While the nonlinear mode-coupling effect directly erases the baryonic features, the accompanied increase in small-scale power due to nonlinear growth will modify the slope of the power spectrum as a function of wavenumber. The resulting change in slope will shift the apparent locations of the baryonic peaks. A careful look at Figure \ref{fig:dlnPlin} reveals that the higher harmonics of the baryonic peaks appear at slightly larger wavenumbers compared to the initial power spectrum. Furthermore this addition of broadband power can misleadingly decrease the contrast of the peaks and thereby overestimate the loss in information by the mode coupling effect on the baryonic features. 

We attempt to correct for the nonlinear growth effect on the broadband shape of the power spectrum by modeling the gradual modification to the broadband shape due to nonlinearity as a smooth function of wavenumber and subtracting off this smooth function from the nonlinear power spectrum to restore the oscillatory portion to its original slope. We fit the nonlinear power spectra in real space to a multiple of the linear power spectrum, $g^2 P_{\rm linear}$, and a $2^{nd}$-order polynomial function $\fL=c_0+c_1\;k+c_2\;k^2$ (with $g^2$ and the $c_i$ as constants) and then subtract the smooth function $\fL$ from the nonlinear power spectra before calculating $d \ln P/d \ln k$. This process increases the fractional variation in $P$ by decreasing the overall amplitude. In other words, noting that $d \ln P/d \ln k =(1/P) \times (dP/d \ln k$), we have decreased $P$ but not substantially changed its derivative.
Increasing the degree of the function $\fL$ up to $3^{rd}$-order does not have a sizable effect on $d\ln P/d\ln k$ on large scales. As required, the resulting function $\fL$ is nearly constant on large scales in all cases. The returned values of $g^2$ are not consistent with linear growth rate at lower redshift because the fitting process tends to decrease $g^2$ to account for the erasure of baryonic features. Since we do not want to over-amplify the baryonic features by oversubtracting beyond a true nonlinear power, as a sanity check, we calculate and subtract $\fL$ by fixing $g^2$ to the analytic linear growth rate. This reduces the contrast of resulting baryonic features in $d\ln P/d\ln k$ at $z=0.3$ but only by a small amount.

The left panel of Figure \ref{fig:dlnPlinminus} shows the resulting nonlinear $d\ln P/d\ln k$ in real space (black lines) in comparison to the linear power spectrum at $z=49$ (gray lines). The agreement between the power at $z=3$ and the linear power spectrum is excellent. We also recover larger contrasts in $z=0.3$ and $z=1$ cases after the broadband shape is restored. Our result shows that it is possible to trace baryonic features up to $k\sim 0.3\ihMpc$ at $z=1$ and $k\sim 0.2 \ihMpc$ at $z=0.3$. The results suggest that the performance of baryonic features as a standard ruler will be diminished due to the decreased contrast at $k < \kmax$, but the features preserved beyond $\kmax$ will tend to compensate for the reduction. 

In short, we find that correcting for the nonlinear effects on the broadband shape with a smooth function helps us rescue baryonic features by some degree, implying that the nonlinear effects are relatively smooth in power. We interpret the amount we cannot recover as having been lost to mode coupling effects.

For the redshift-space power spectrum, we first correct for the finger-of-God suppression. We fit the nonlinear power spectrum in redshift space to the counterpart in real-space by a multiplicative smooth function in the form of $F_{\rm fog}=1/(k^m \sigma^m+1)^{1/m}$ where $\sigma$ and $m$ are fitting parameters. We find $m \sim 2.3-2.95$. After dividing by $F_{\rm fog}$ we calculate and subtract $\fL$ to match the restored redshift-space power spectrum to the linear power spectrum. Although $F_{\rm fog}$ may not necessarily match the conventional form of an exponential finger-of-God suppression, the function behaves properly at the limits of small and large wavenumber and is capable of characterizing the difference between the real-space and redshift-space power from our N-body results. The right panel of Figure \ref{fig:dlnPlinminus} shows the resulting $d\ln P/d\ln k$ in redshift space (black lines) in comparison to the linear power spectrum at z=49 in real space (gray lines). The restored redshift-space power spectra show slightly larger contrasts in baryonic features at lower redshift relative to the uncorrected ones (lower right panel of Figure \ref{fig:dlnPlin}). From the comparison between the real-space and redshift-space $d\ln P/d\ln k$ (Figure \ref{fig:dlnPlinminus}), the redshift-space power reasonably traces the details in the real-space power on linear and quasilinear scales although the contrast of the baryonic features degrades in redshift space.

In summary, the real-space (redshift-space) matter power spectrum traces baryonic features up to $k\sim 0.4$ $(0.3)\ihMpc$ at $z=3$. At $z=1$, $\kmax$ of $0.19\ihMpc$ seems a reasonable choice for the linear approximation in real space considering the features preserved on even smaller scales, but a lenient standard in redshift space. At $z=0.3$, $\kmax\sim 0.11\ihMpc$ is conservative both in real space and redshift space. Adopting a slightly larger $\kmax$ appears justified at this redshift. 

\begin{figure*}[t]
\epsscale{2}
\plottwo{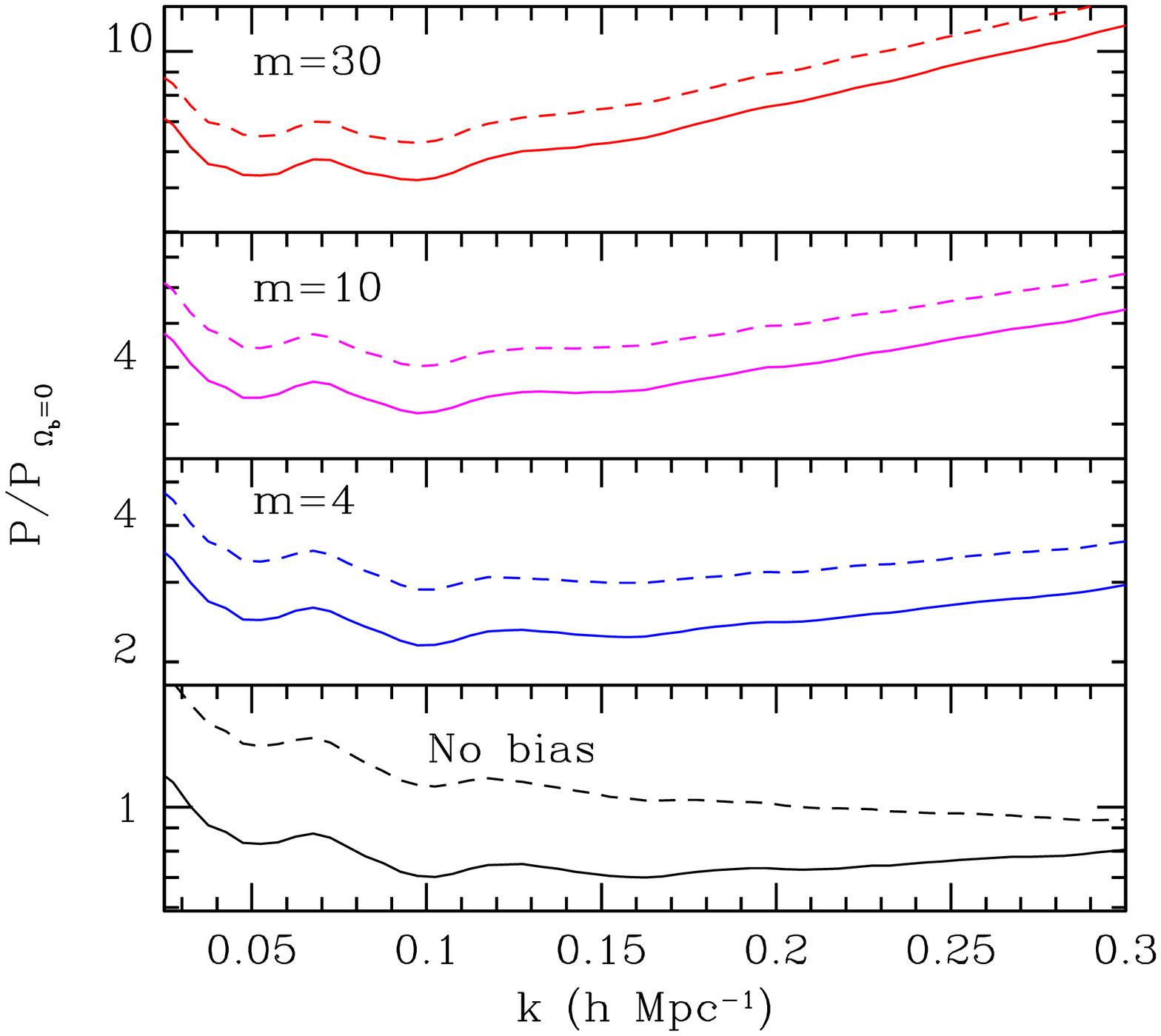}{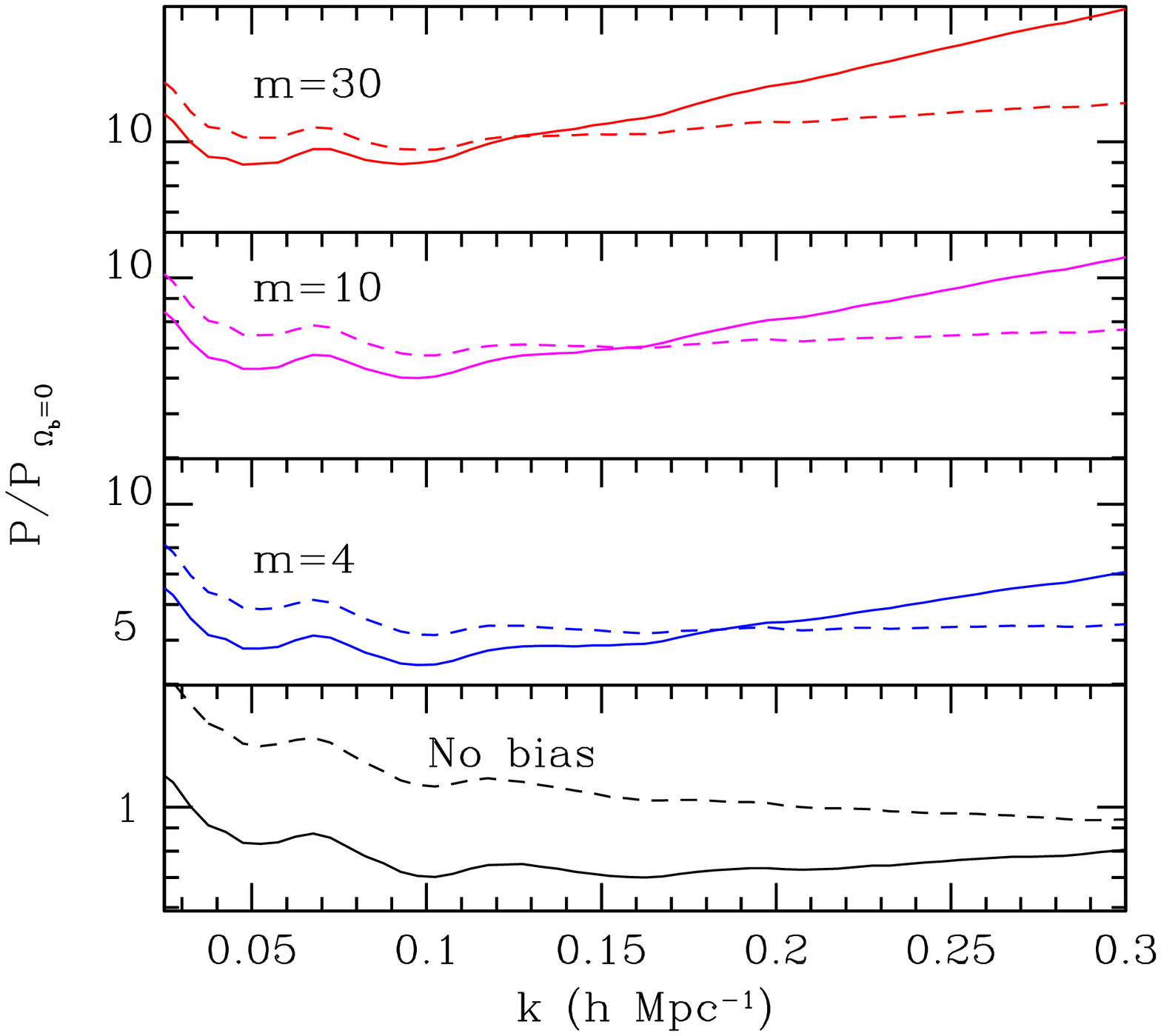}
\caption{Biased power spectra at $z=1$ divided by a zero-baryon power spectrum. The left panel is for the number-weighted cases (NUM), and the right panel is for the mass-weighted cases (MASS). The label `$m$' denotes the minimum group multiplicity for selecting halos. For the NUM cases, $b \sim 1.7$, 2, and 2.5, and for the MASS cases, $b \sim 2.4$, 2.7, and 3.1 as $m$ increases. The solid lines are for the real-space power, and the dashed lines are for the redshift-space power. }\label{fig:T0bz1}
\epsscale{1}
\end{figure*}

\section{The effect of bias}\label{sec:bias}

\subsection{Anomalous power}\label{subsec:shotnoise}
In galaxy redshift surveys, we do not directly observe the real-space matter power spectrum but instead observe the distribution of biased tracers of matter in redshift space. The assumptions of local bias and Gaussian statistics for the density fields lead to a scale-independent bias on large scales for the correlation function \citep{Coles93,Sch98,Meiksin99,Col99}. Any excess small-scale correlation from biasing will appear as an additional constant term in the biased power spectrum on large scales. This holds even when the matter density field is nonlinear \citep{Sch98,Col99,Sel00}. The bias on large scales is thus scale-independent up to this constant term. On smaller scales, the bias will generically deviate from the simple approximation of scale independence. We use the term `nonlinear bias effect' in this paper to designate any deviation from a simple multiplicative bias in the power spectrum.

As the biased tracers such as galaxies are rare objects compared to the underlying matter, the biased power spectrum is subject to a shot noise. Conventionally, one writes the shot noise as a white noise ($P \sim$ constant) equal to the inverse of the number density of particles. 
However, the effects of the limited number of particles can be more complicated than a simple Poisson noise, such as in biasing schemes where halos are not equally weighted or, more enigmatically, the discreteness effects in the process of finding halos. For example, the fact that a halo finder cannot identify two halos arbitrarily close together means that the shot noise will not be white. Therefore it is not straightforward to identify the shot noise with the conventional white noise based on an inverse number density. 
Rather than singling out the shot noise term, we hereafter group this term with the nonlinear bias term without distinguishing one from the other. Both terms are easily approximated as an additive constant on large scales. Hereafter, `anomalous power' is used to refer to a combination of these two terms. We find that the fractional level of the anomalous power compared to the linearly biased power is important to track. We hereafter will describe this {\it fractional} level as `larger' or `smaller' anomalous power. 

The anomalous power will contribute additional power above the linearly biased power spectrum, therefore increasing the statistical variance of the underlying features. The nonlinear bias effect in the anomalous power may also induce a mixture of information from different Fourier modes and so erase features.

\begin{figure*}[t]
\epsscale{2}
\plottwo{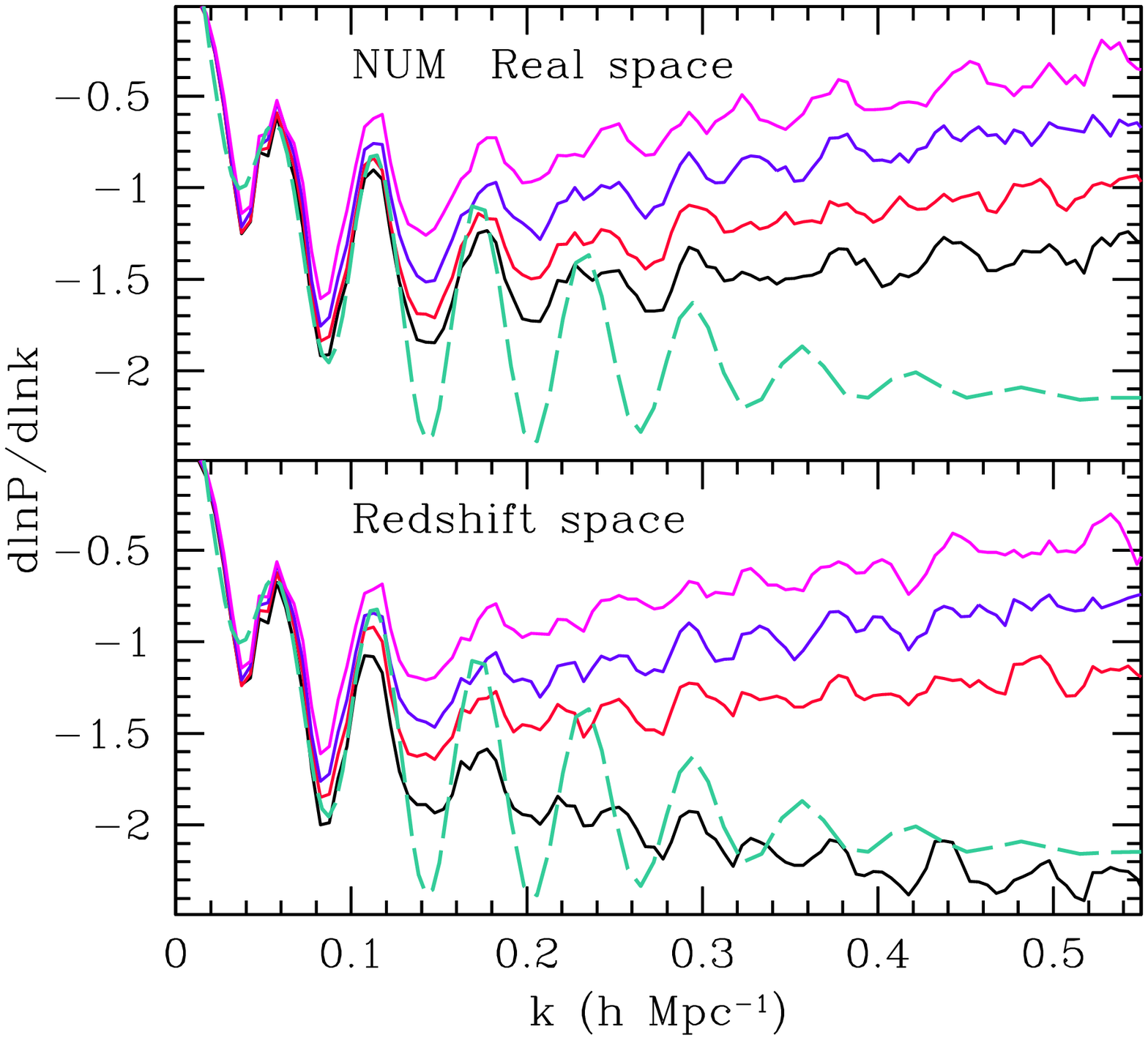}{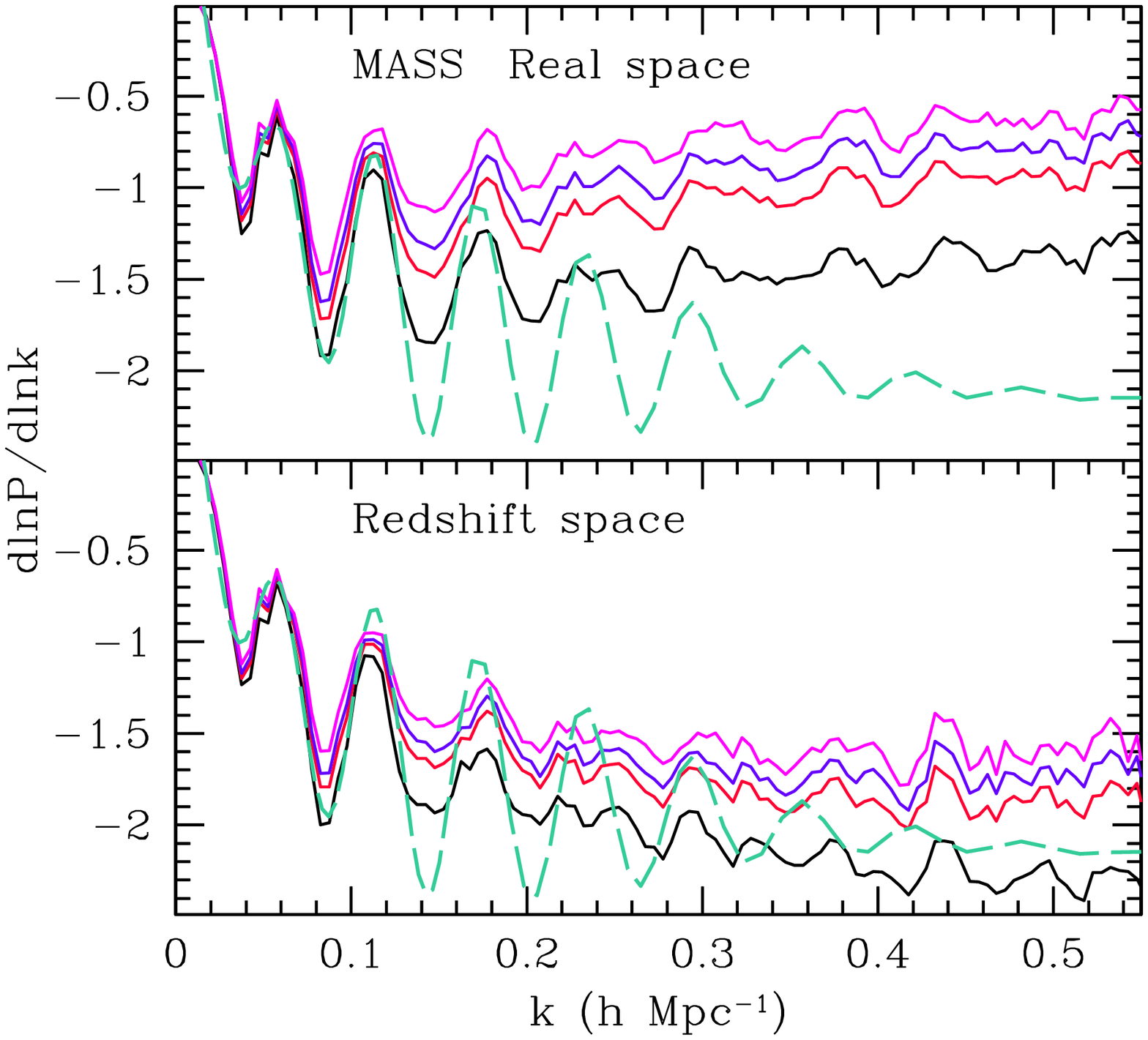}
\caption{$d\ln P/d\ln k$ from the biased power spectra at $z=1$ in real space (upper panels) and in redshift space (lower panels). The left panel is for the number-weighted cases (NUM), and the right panel is for the mass-weighted cases (MASS). Green dashed line: the input power spectrum, black: the nonlinear matter power spectrum at $z=1$, red: biased with $m=4$, blue: biased with $m=10$ and violet: biased with $m=30$.}
\label{fig:dlnPz1}
\epsscale{1}
\end{figure*}

\subsection{Bias schemes}\label{subsec:biasschemes}
In generating biased tracers of the matter, we do not attempt to reproduce realistic galaxy populations but instead to reflect an interesting range of models of galaxy populations using simple deterministic halo-based biasing schemes. 
The current halo occupation distribution (HOD) models suggest a single galaxy per halo for low-mass halos above a mass threshold and an additional power-law mean occupation for more massive halos while the details vary in different studies of galaxy populations \citep{Berlind03,Krav04,Zehavi04}. 
We reduce the complexity by decomposing the HOD models into the two extreme bounding cases: one in which halos above a mass threshold host a single galaxy and one in which halos above a mass threshold have mean occupation as linear to the halo mass (i.e., the power-law index of unity). To create different amplitudes of bias, we apply various minimum group multiplicities (i.e., halo mass-thresholds) for both cases. 
Various superpositions of these trial cases then can comprise more complex HOD models. Note that the resulting populations themselves already represent superpositions of different-mass halos. 
Therefore our biasing schemes will be sufficient for examining the robustness of the baryonic features in various galaxy populations because, if baryonic oscillations are found to survive in both of the extreme cases, it seems unlike that a mixture would fair worse. This will be further justified once our results can show that different bias models indeed extrapolate to linear biasing on large scales. 

We use the friends-of-friends method \citep{Davis85} and identify halos by adopting a linking length of 0.6 $\hMpc$ and minimum group multiplicities of 4, 10, or 30 particles. We assign zero galaxy density for the regions below this threshold. For the regions identified as halos with a given threshold, we assign galaxies with two schemes:

1. Number Weighted (NUM), where we assign one galaxy per halo, using the center-of-mass position and velocity. In this case, the number of galaxies does not follow the mass of halos, and the information on the virialized motions within each halo (the finger-of-God effect) is lost.

2  Mass Weighted (MASS), where we retain the velocity and density structure of the halo by assigning one galaxy per simulation particle. This way, the number of galaxies follows the mass of the halo, and the finger-of-God effect is preserved.

The minimum group multiplicity, $m$, of 4 and 10 are small compared to $20- 30$ particles usually desired for robust halo identification \citep{Som00,Jenk01}. However, the low multiplicity halos are still tracing overdense regions, albeit in a more stochastic manner. This randomness would not be favorable for recovering acoustic oscillations, implying that our results are conservative.

At lower redshifts, both methods are able to provide us a sufficient range of bias values of our interest. However, at $z=3$, the mass resolution of our simulation is too low to find halos with masses low enough to yield mild values of biases (i.e. $b\sim 3$). For this reason, we use an additional bias scheme at this redshift, where the density of the tracers approximately follows the matter density squared (hereafter, $\Ro$). 
In this biasing scheme, each particle is weighted by the average density with a $2\hMpc$ spline-smoothing kernel centered on the particle. We implement this with the SMOOTH code\footnote{http://www-hpcc.astro.washington.edu/tools/}. From the distribution of the weighted particles, we calculate the density in each mesh in real space and redshift space. The finger-of-God effect is preserved in redshift space in this scheme.

We first discuss the biased power spectra at $z=1$ and $z=0.3$ because the two share common bias schemes. We then present the $z=3$ results.

\subsection{Bias effects at $z=1$} 

Figure \ref{fig:T0bz1} shows power spectra of biased tracers divided by a zero-baryon power spectrum in real space and redshift space. The left and right panels show number-weighted cases (NUM) and mass-weighted cases (MASS), respectively, with the minimum group multiplicity $m$, of 4, 10, and 30.

NUM cases with $m=4$, 10, and 30 produce biased tracers with $b \sim 1.7$, 2, and 2.5 where $b$ is the ratio of the biased power spectra to the matter power spectra on large scales. 
The mass-weighted tracers (MASS) in the right panel of Figure \ref{fig:T0bz1} show larger biases than number-weighted cases for the same $m$: $b \sim 2.4$, 2.7, and 3.1. This is because the mass-weighted cases give more weight to the high-mass halos with larger biases while the number-weighted cases are dominated by halos close to the mass threshold. 

In both cases, the relative anomalous power increases as the amplitude of bias increases. 
For example, the $m=4$ cases show little anomalous power, and the $m=10$ and $m=30$ cases show larger but still mild anomalous power for $k< \kmax$ ($0.19 \ihMpc$). When different bias models with the same multiplicity are compared, the MASS cases have slightly larger anomalous power up to $k\sim 0.3\ihMpc$. When different bias models with a similar value of bias are compared (e.g., $m=4$ of MASS and $m=30$ of NUM), the MASS case exhibits smaller anomalous power up to $k \sim 0.3\ihMpc$ than the NUM case. MASS cases with all galaxies placed at the center of a halo showed the same trend, which means that the trend is not due to the effect of the halo density-profile. Thus, the bias scheme for the NUM cases introduces larger anomalous power, be it from the shot noise or from the nonlinear bias effect. One interesting question will be whether the difference in anomalous power directly relates to the erasure of baryonic features.

The upper panels in Figure \ref{fig:dlnPz1} show $d\ln P/d\ln k$ of biased power in real space (solid lines) in comparison to the input power spectrum (dashed lines). In the figure, all biased power spectra in real space preserve oscillatory features at least up to $k\sim 0.2 \ihMpc$ while the contrast appears decreasing due to the nonlinear bias effect as bias increases. As before (\S~\ref{subsec:nonlinminus}), adding a smooth component due to a shot noise or nonlinear bias to the power spectrum will decrease the contrast in baryonic features. Subtracting off this smooth component will help to recover some of the baryonic features. We will revisit this in \S~\ref{subsec:nonbiasminus}.

In redshift space (dashed lines in Figure  \ref{fig:T0bz1}), the power spectra of the MASS cases clearly show the finger-of-God suppression as $k$ increases. The finger-of-God suppression is mostly removed in the spectra of the NUM cases, as is to be expected from the methods of biasing, and they show a mild remnant suppression in power with respect to the Kaiser formula\footnote{Kaiser formula in our context means $(1+2\beta (k)/3 + \beta^2 (k)/5)$ where $\beta (k)=\Om^{0.6}(z)/b(k)$, and $b(k)$ is calculated from the ratio between the biased power spectra and the matter power spectra.} as $k$ increases. The lower panels of Figure \ref{fig:dlnPz1} show that baryonic oscillations are smeared more in redshift space than in real space not only in MASS cases but also in NUM cases despite the suppressed finger-of-God effect in the latter case. Again, until we remove the bias effect on the broadband shape, it is hard to determine the degree of erasure.

\subsection{Bias effects at $z=1$ with the broadband shape restored} \label{subsec:nonbiasminus}
We next subtract the anomalous power from the biased power spectrum to restore the broadband shape and eliminate the superficial decrease in contrast of the baryonic features. The restoring process we adopt is intended to assess the optimal amount of information on baryonic features available from the biased tracers. In real galaxy redshift surveys, this unbiasing would be determined simultaneously in the parameter estimation, which is an additional complication.

\begin{figure*}[t]
\epsscale{2}
\plottwo{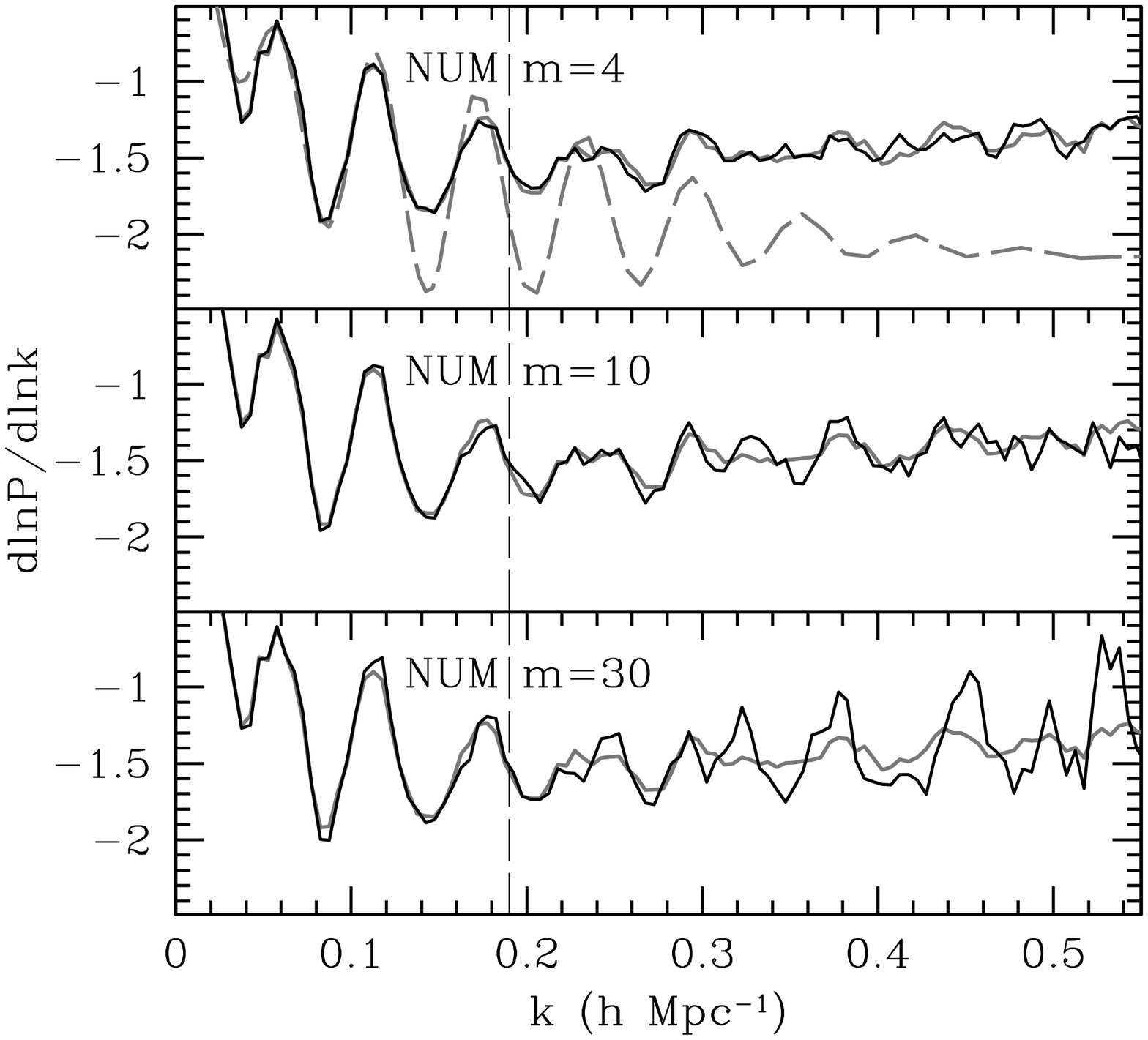}{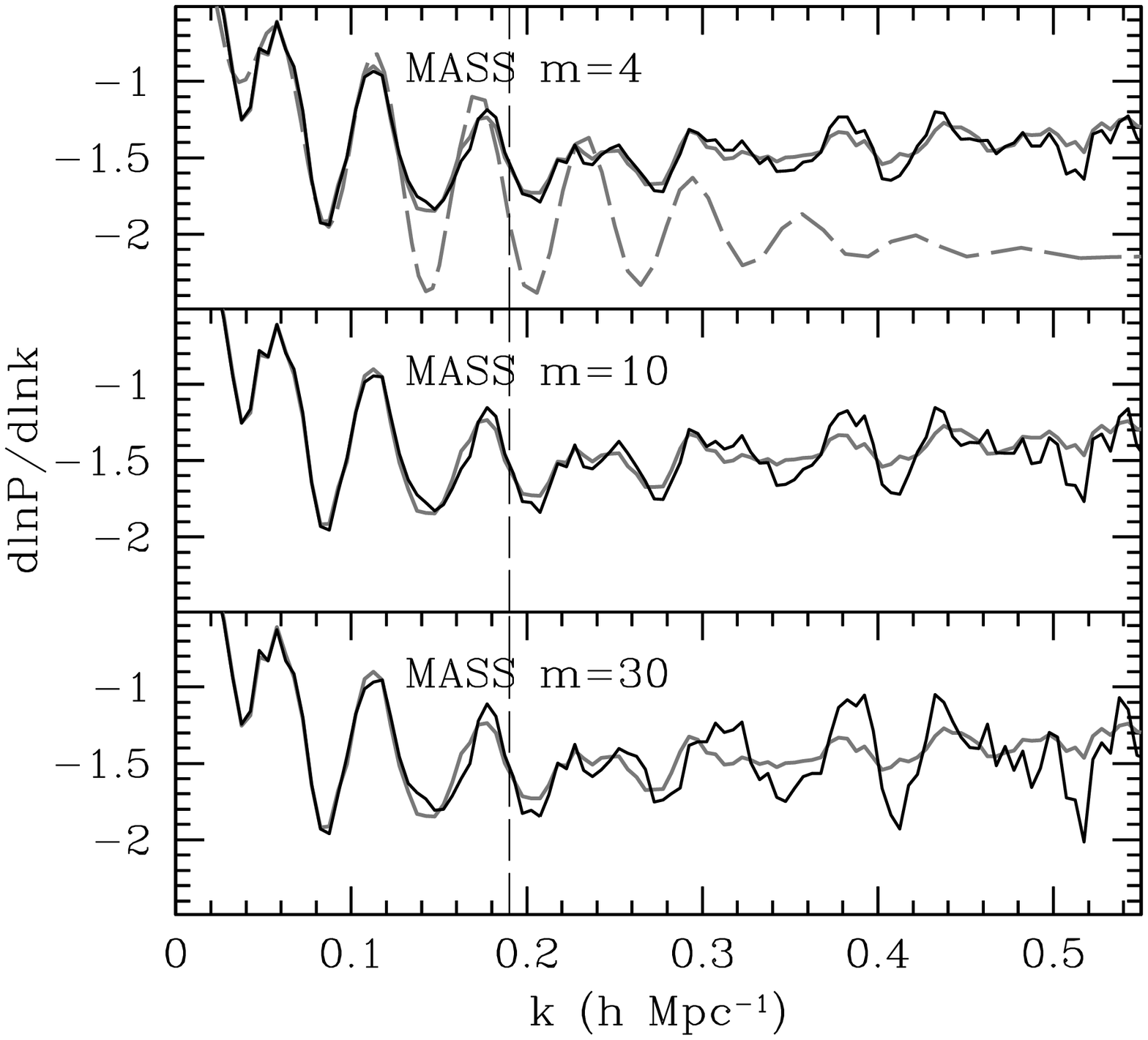}
\caption{$d\ln P/d\ln k$ from $P_{\rm biased}-\fNL(c_0,k,k^2)$ at $z=1$ in real space. The left panel is for the number-weighted cases (NUM), and the right panel is for the mass-weighted cases (MASS). The label $m$ denotes the minimum group multiplicity for selecting halos. Gray dashed line: the input power spectrum, gray solid line: the nonlinear matter power spectrum at $z=1$, black solid lines: the biased power spectra at $z=1$. The vertical dashed line denotes $\kmax=0.19 \ihMpc$. One sees that the biased power spectra follow the features in the underlying matter power spectrum fairly well on linear and quasilinear scales once the broadband shape is restored.}
\label{fig:dlnPdlnkminusz1}
\epsscale{1}
\end{figure*}
\begin{figure*}[t]
\epsscale{2}
\plottwo{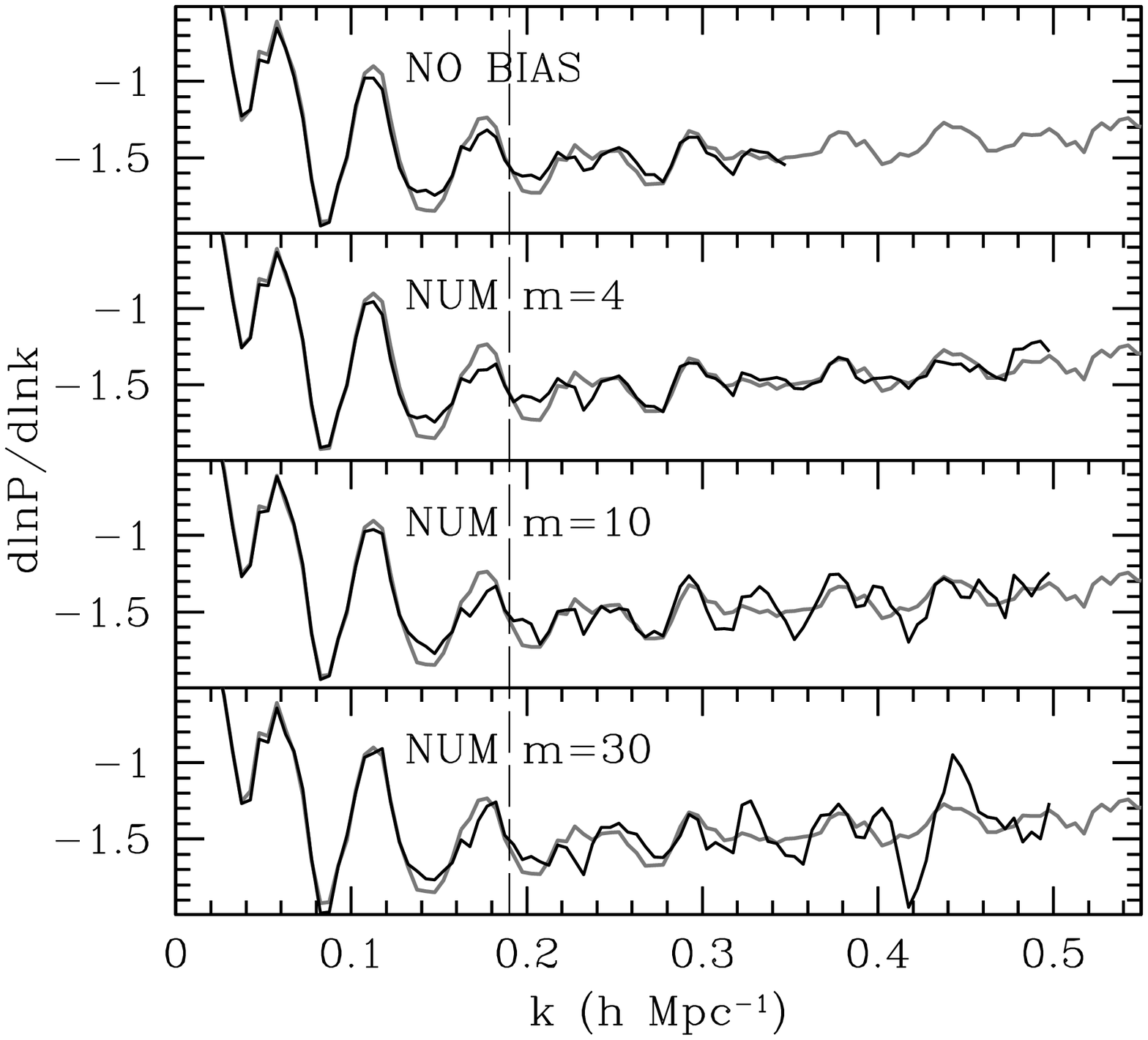}{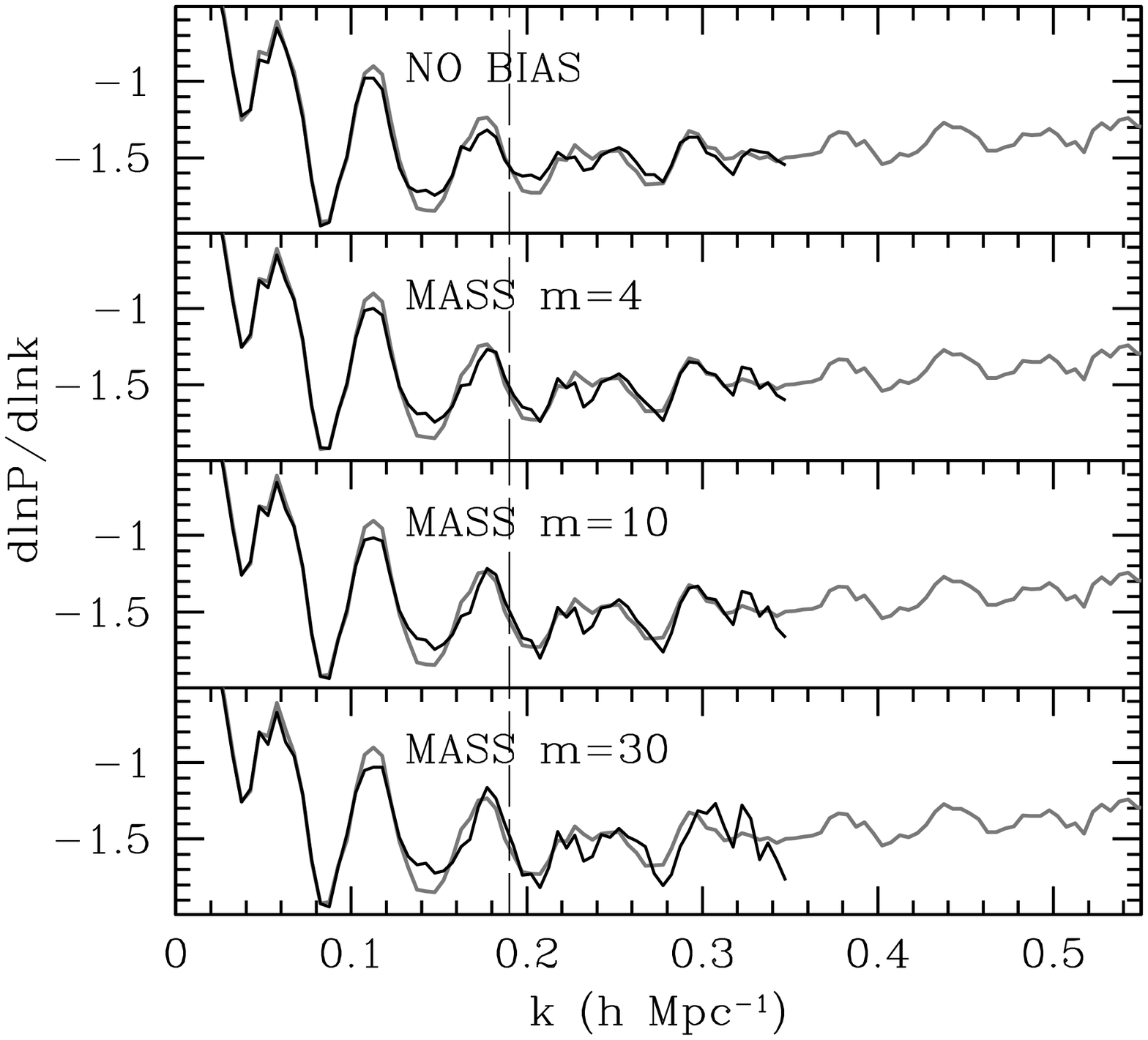}
\caption{$d\ln P/d\ln k$ from $P_{\rm biased}-\fNL(c_0,k,k^2)$ at $z=1$ in redshift space. Note that the MASS cases are corrected for the finger-of-God suppression beforehand. The left panel is for the number-weighted cases (NUM), and the right panel is for the mass-weighted cases (MASS). Gray: the nonlinear matter power spectrum at $z=1$ in {\it real} space, black: the nonlinear matter power spectrum in redshift space (NO BIAS) and biased power spectra in redshift space. Note that the matter power spectrum in redshift space (NO BIAS) is also corrected with $F_{\rm fog}$ to fit to the matter power in real space. The fitting range ($k < \kfit$) is indicated by the extent of the black lines. The number-weighted cases are fitted to $\kfit=0.5\ihMpc$ but $\kfit=0.35\ihMpc$ elsewhere. The vertical dashed line denotes the value of $\kmax$. The redshift-space biased power spectrum nearly reproduces the baryonic features of the real-space biased power spectrum with mild degradation.}\label{fig:dlnPdlnkminusz1red}
\epsscale{1}
\end{figure*}

We fit the biased power spectra to a multiple of the nonlinear matter power spectrum $b^2 P_{\rm matter}$ at the given redshift plus a polynomial function $f_{\rm NL}=c_0+c_1\;k+c_2\;k^2$. We then subtract the smooth function $\fNL$ that represents the anomalous power from the biased power spectra when calculating $d\ln P/d\ln k$. As the anomalous power from bias is smooth, this will help recover baryonic features. The ranges of wavenumber in real or redshift space is chosen suitably at each redshift so as to span well beyond the linear region to constrain $\fNL$ but short enough so as not to weight the fit too much towards large wavenumbers. Variations in $\fNL$ due to choosing different ranges of wavenumbers or different degrees of the function (up to $3^{rd}$-order) do not show a meaningful impact on $d \ln P/d \ln k$ on large scales. The resulting function $\fNL$ behaves as constant on large scales as required, although on small scales, $\fNL$ for the MASS cases show a slow roll-over due to the extended halo profiles compared to the NUM cases.

Figure \ref{fig:dlnPdlnkminusz1} shows the resulting derivatives at $z=1$ after anomalous power $\fNL$ is subtracted (black lines) in comparison to the nonlinear (solid gray lines) and input matter power spectra (dashed gray lines). For $k< \kmax$, the agreement between the biased power spectra and the nonlinear matter power spectrum is excellent regardless of the different biasing schemes. Beyond $\kmax$, we observe small variations depending on bias schemes. 
The variations appear related to the amount of anomalous power as this contributes additional power above the underlying baryonic features, increasing the statistical noise. Within the same bias models, higher mass thresholds and hence larger biases yield noisier derivatives, as would be expected from the increase in anomalous power. If we compare two bias models with similar bias values, $m=4$ of MASS and $m=30$ of NUM, then the baryonic features in $m=30$ of NUM appear noisier as it has a larger anomalous power. However, when those with the same group multiplicity are compared between different bias models, the baryonic features in the MASS cases look no worse than those in the NUM cases even with the larger bias and the slightly larger anomalous power for $k\lesssim 0.3\ihMpc$.

Despite slight variations depending on bias schemes, the baryonic features in general have not been obviously damaged by the biasing process; the biased power spectrum closely follows the details in the underlying nonlinear matter power spectrum over a broad range beyond $\kmax$. This is different from the effect of nonlinear gravity shown in \S~\ref{subsec:nonlinminus} where we saw the effect of mode coupling. This supports the hypothesis that the anomalous power, whether due to shot noise or nonlinear bias, is a smooth function of wavenumber. The contrasts of the baryonic features appear slightly larger in some of the biased power spectra relative to the matter power spectrum, but this is likely due simply to the increased noise.

  For the redshift-space power spectrum of the MASS cases, we apply a similar fitting process to the one in \S~\ref{subsec:nonlinminus} to restore the broadband shape both from the nonlinear bias and nonlinear redshift distortions. First we correct for the finger-of-God effect with a multiplicative function $F_{\rm fog}$ to match the biased power in redshift space to the biased power in real space. We then calculate and subtract $\fNL$ to remove the anomalous power. For the redshift-space power spectrum of the NUM cases, we do not correct for the finger-of-God suppression despite the slight deviation from the Kaiser formula. We calculate and subtract an additive $\fNL$ from the redshift-space power in this case. 

 Figure \ref{fig:dlnPdlnkminusz1red} shows the resulting derivatives at $z=1$ in redshift space (black lines) in comparison to the nonlinear matter power spectrum in real space (gray lines). In redshift space, the contrast of the last feature before $k=0.2\ihMpc$ is smaller than in real space but still in good agreement. Beyond $\kmax$, we see the traces of baryonic features although they look noisier than in real space. Again, the nonlinear effect of redshift distortions on baryonic features in the NUM cases are no better than that in the MASS cases even though the virialized motions within the halos are suppressed in NUM cases, and this probably is related to the nonlinear effect on the velocity fields on large scales \citep{Sco04}.

\subsection{Bias effects at $z=0.3$} 
\begin{figure*}
\epsscale{2}
\plottwo{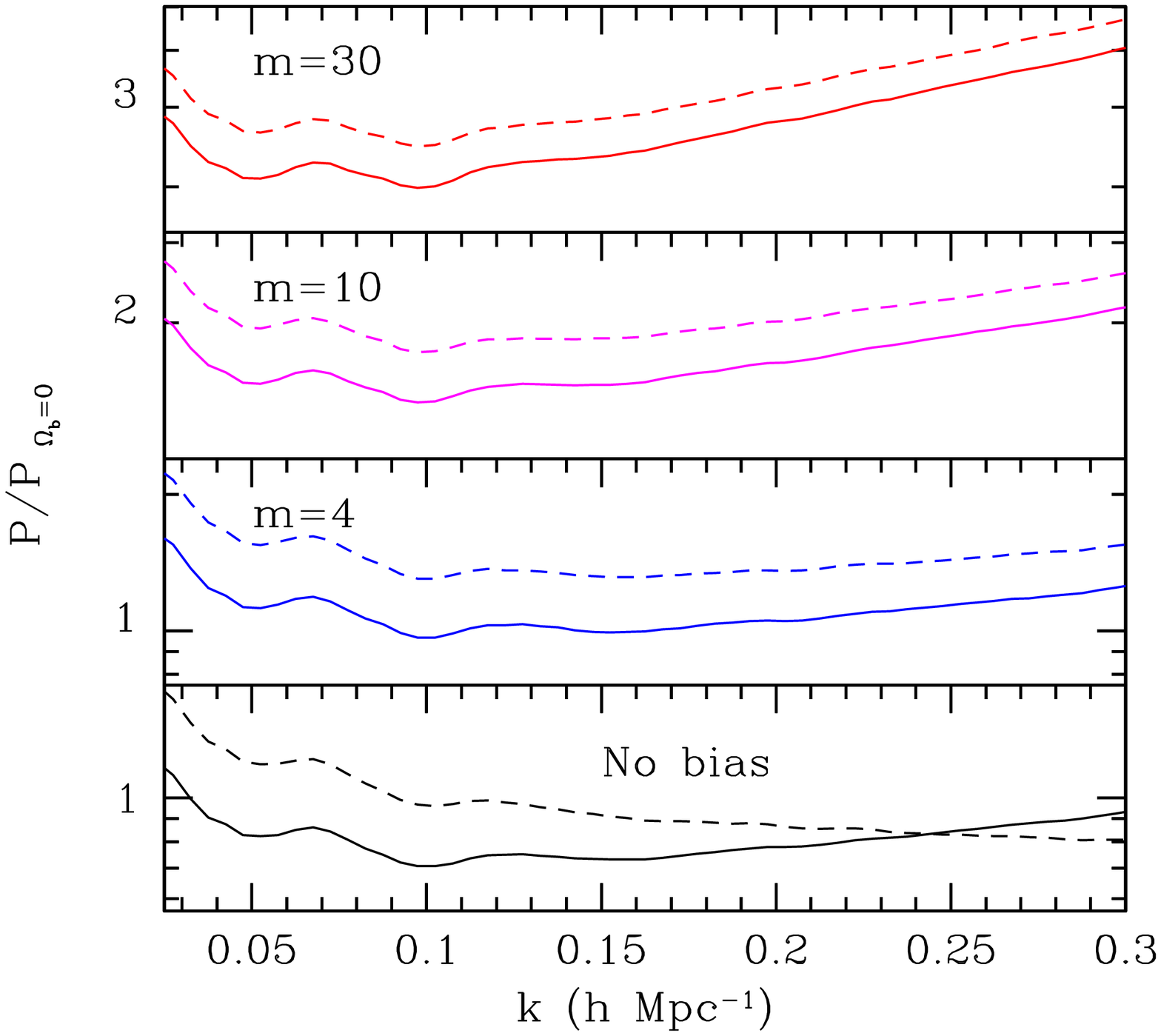}{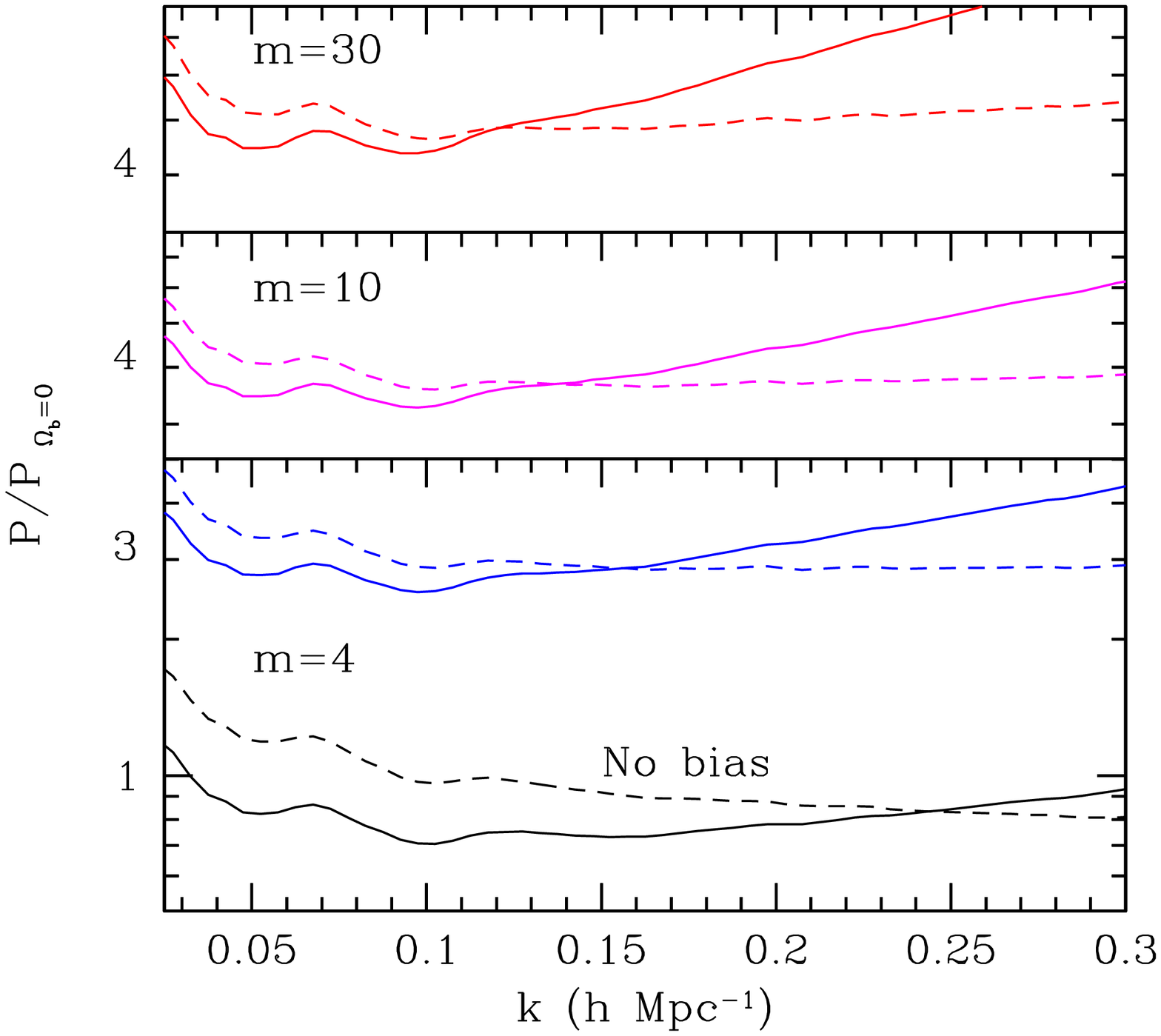}
\caption{Biased power spectra at $z=0.3$ divided by a zero-baryon power spectrum. The left panel is for the number-weighted cases (NUM), and the right panel is for the mass-weighted cases (MASS). The label $m$ denotes the minimum group multiplicity. For the NUM cases, $b \sim 1.2$, 1.3, and 1.6, and for the MASS cases, $b \sim 1.8$, 2, and 2.3 as $m$ increases. The solid lines are for the real-space power, and the dashed lines are for the redshift-space power. }
\label{fig:T0bz0.3}
\epsscale{1}
\end{figure*}

\begin{figure*}
\epsscale{2}
\plottwo{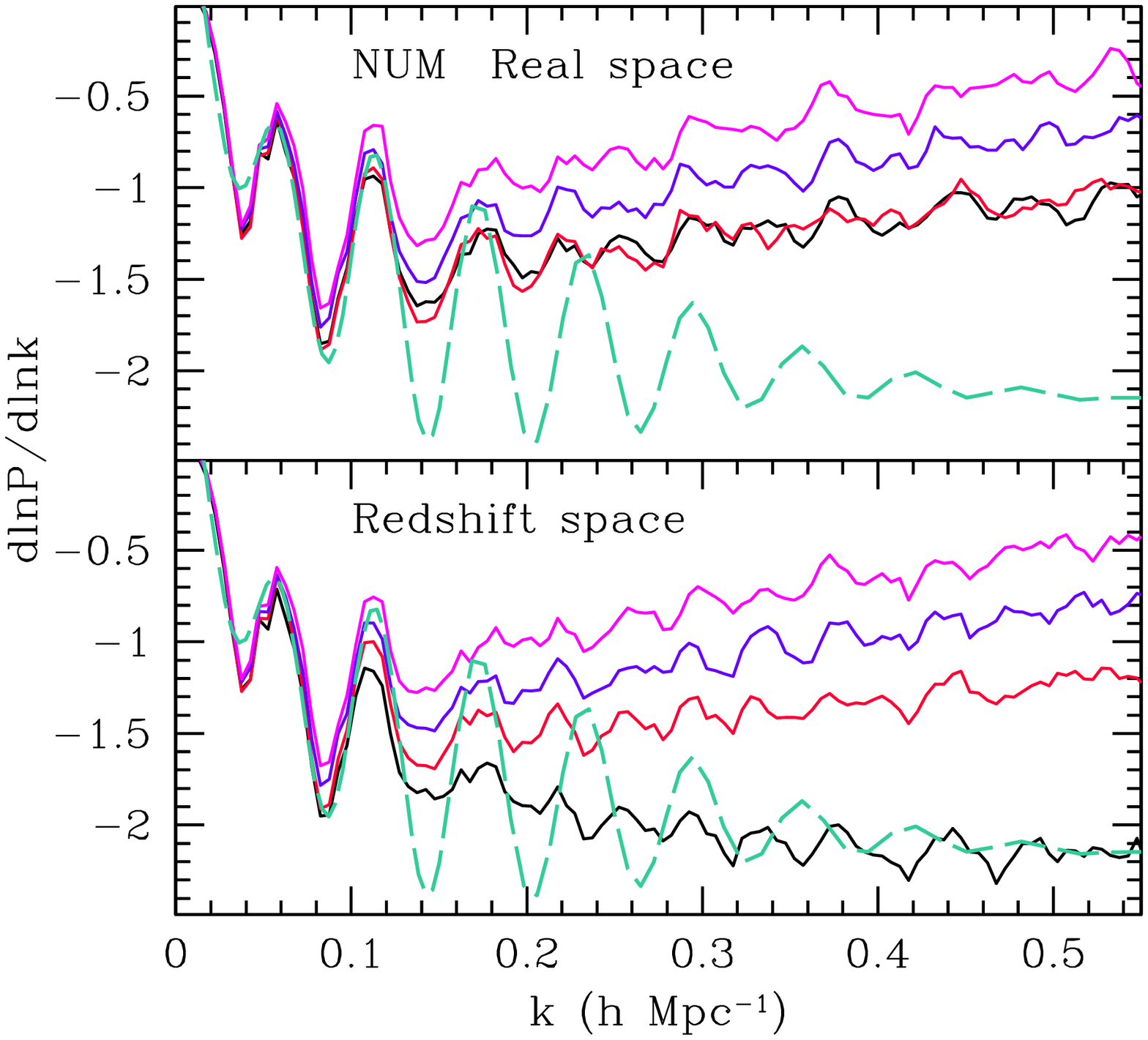}{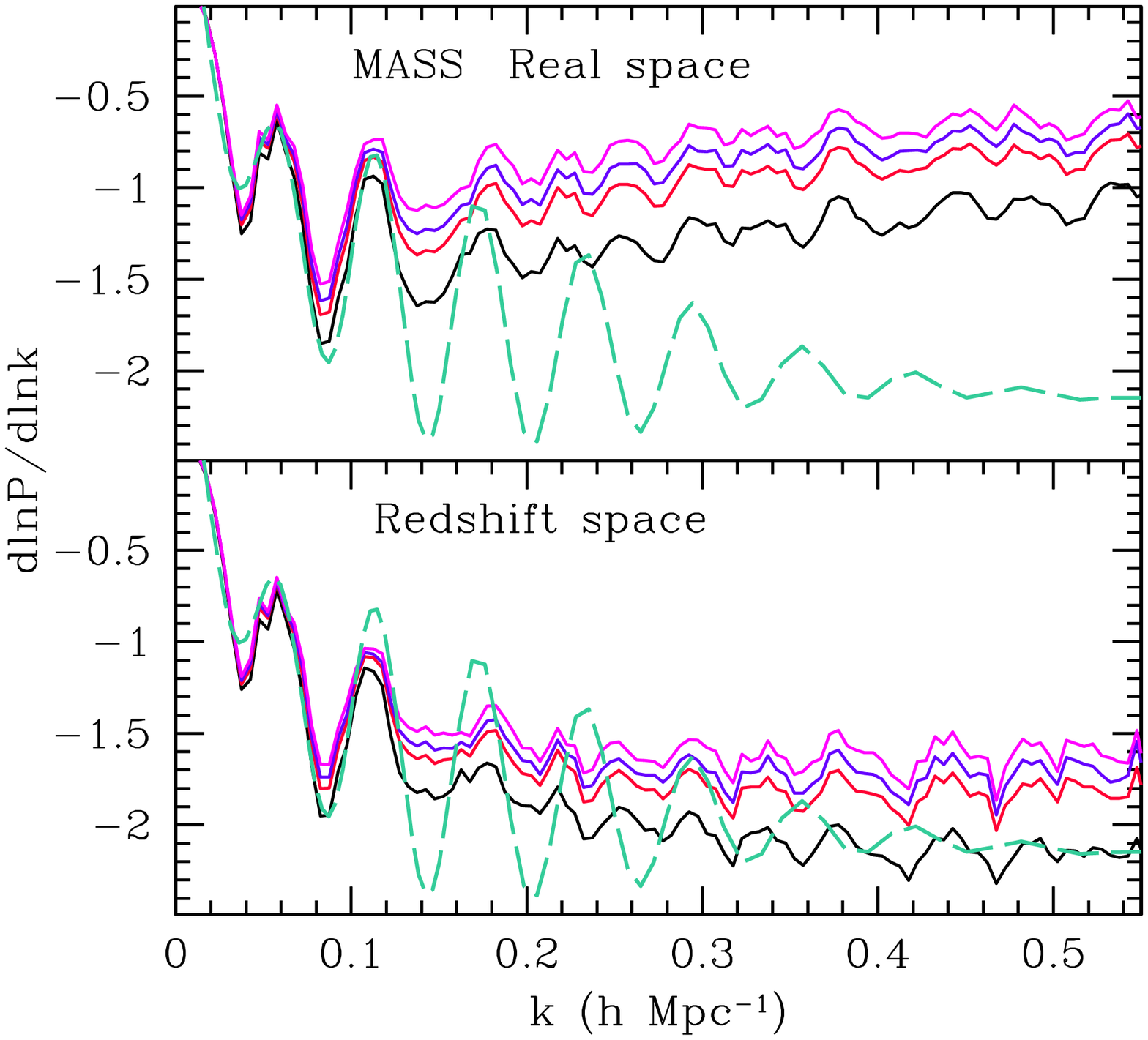}
\caption{$d\ln P/d\ln k$ from the biased power spectra at $z=0.3$ in real space (upper panels) and redshift space (lower panels). The left panel is for the number-weighted cases (NUM), and the right panel is for the mass-weighted cases (MASS). Green dashed line: the input power spectrum, black: the nonlinear matter power spectrum at $z=0.3$, red: biased with $m=4$, blue: biased with $m=10$ and violet: biased with $m=30$.}
\label{fig:dlnPz0.3}
\epsscale{1}
\end{figure*}

We next investigate the effects of bias at lower redshift. Figure \ref{fig:T0bz0.3} shows power spectra of biased tracers divided by a zero-baryon power spectrum at $z=0.3$. The NUM cases (left panel) generate tracers with $b\sim 1.2$, 1.3, and 1.6, and the MASS cases (right panel) generate $b\sim 1.8$, 2, and 2.3. At this redshift, we are particularly interested in tracers with $b \sim 2$, which corresponds to the luminous red galaxy sample (LRG) of Sloan Digital Sky Survey (SDSS). This corresponds to $m=10$ in the MASS cases although anomalous power in the $m=10$ case is half the inverse of the number density of galaxies in the LRG sample, suggesting that the $m=10$ bias model is not exactly right. The bias and relative anomalous power is small up to $k \sim 0.15 \ihMpc$ in all cases relative to $z=1$. Recovering baryonic features beyond $\kmax$ ($0.11 \ihMpc$) will be possible both in real space and in redshift space from Figure \ref{fig:dlnPz0.3}. 


\begin{figure*}
\epsscale{2}
\plottwo{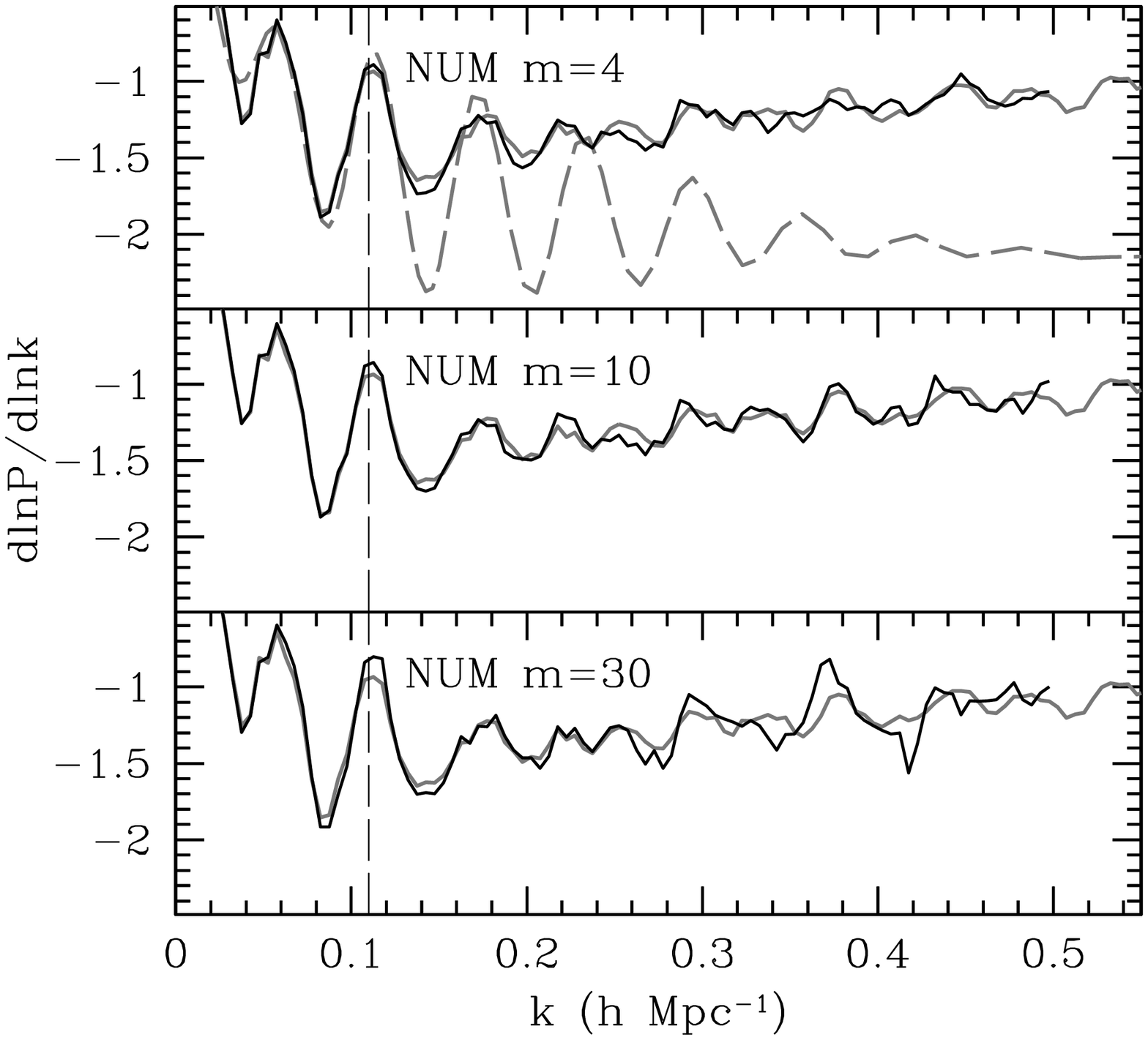}{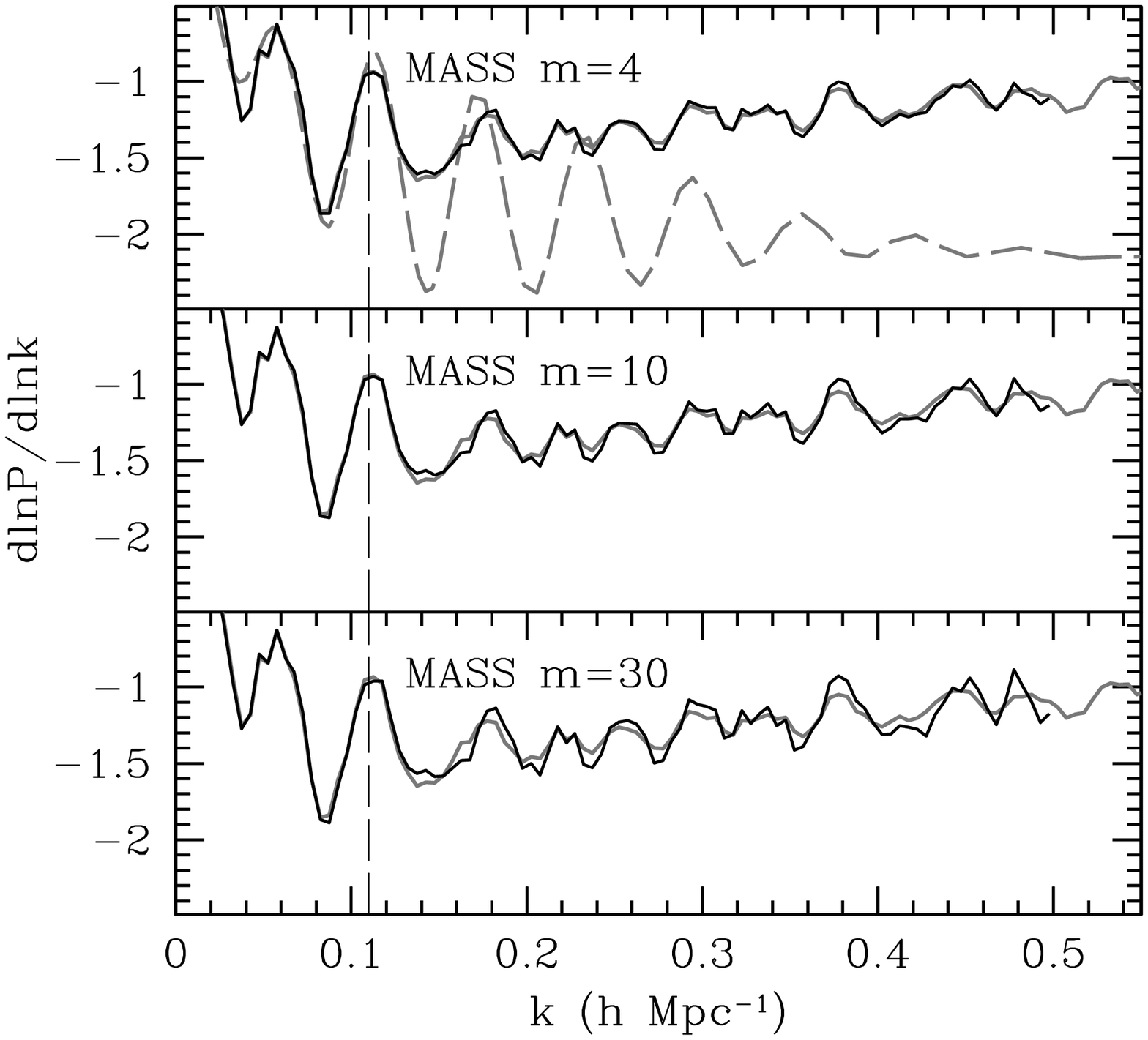}
\caption{$d\ln P/d\ln k$ from $P_{\rm biased}-\fNL(c_0,k,k^2)$ at $z=0.3$ in real space.  The left panel is for the number-weighted cases (NUM), and the right panel is for the mass-weighted cases (MASS). Gray dashed line: the input power spectrum, gray solid line: the nonlinear matter power spectrum at $z=0.3$, black solid lines: the biased power spectra. Fitting to $\kfit=0.5\ihMpc$. The vertical dashed line denotes $\kmax = 0.11 \ihMpc$.  }
\label{fig:dlnPdlnkminusz0.3}
\epsscale{1}
\end{figure*}

\begin{figure*}
\epsscale{2}
\plottwo{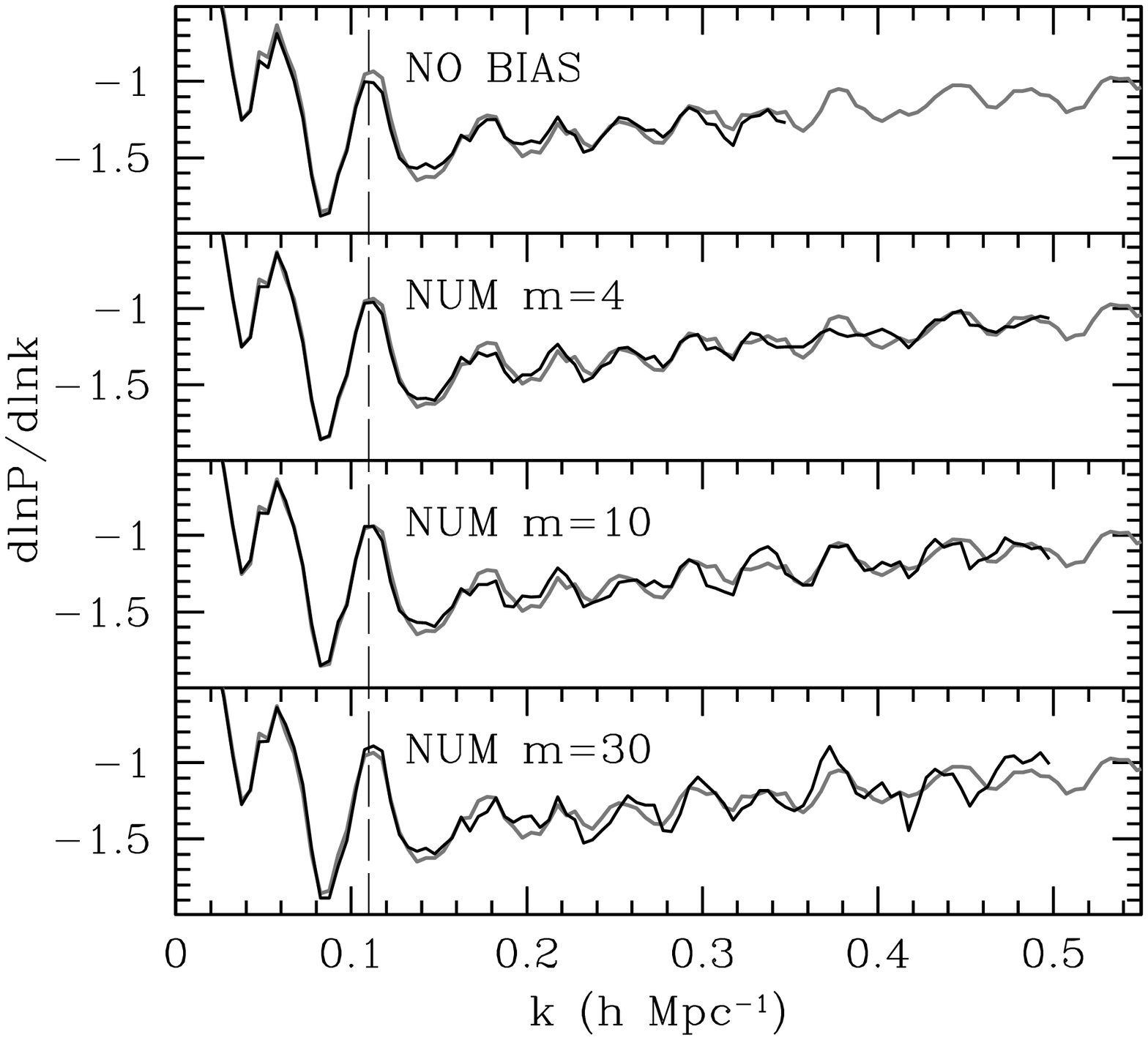}{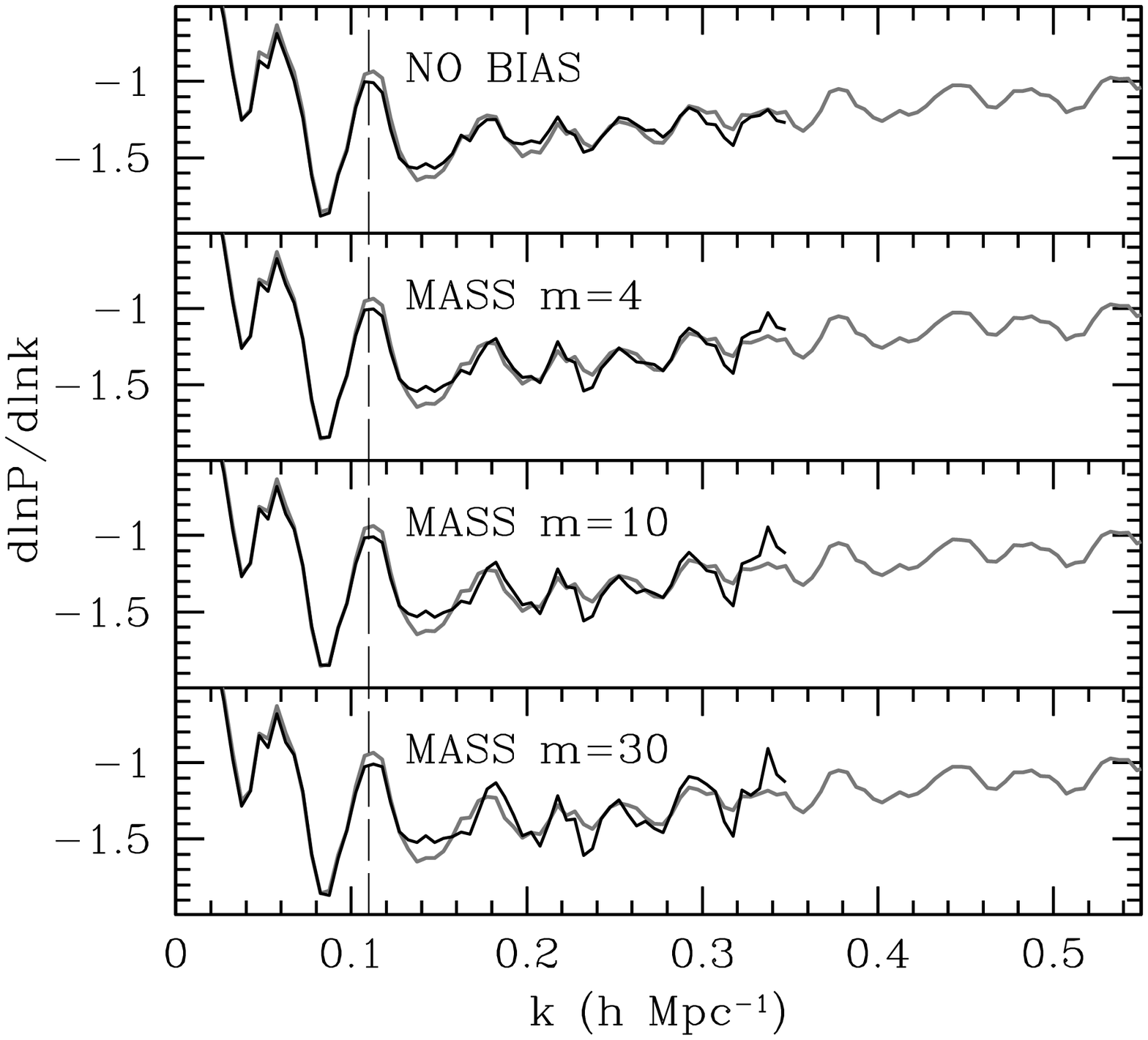}
\caption{$d\ln P/d\ln k$ from $P_{\rm biased}-\fNL(c_0,k,k^2)$ at $z=0.3$ in redshift space. Note that the MASS cases are corrected for the finger-of-God suppression beforehand. The left panel is for the number-weighted cases (NUM), and the right panel is for the mass-weighted cases (MASS). Gray: the nonlinear matter power spectrum in {\it real} space, black: the nonlinear matter power spectrum in redshift space (NO BIAS) and biased power spectra in redshift space. The fitting range ($k < \kfit$) is indicated by the extent of the black lines.  For the number-weighted cases (NUM), $\kfit=0.5\ihMpc$ but $\kfit=0.35\ihMpc$ elsewhere. The vertical dashed line denotes the value of $\kmax$.}
\label{fig:dlnPdlnkminusz0.3red}
\epsscale{1}
\end{figure*}

Figure \ref{fig:dlnPdlnkminusz0.3} shows the derivatives in real space after a corresponding smooth function $\fNL$ is subtracted. From the figure, the baryonic features in biased power spectra trace those in the matter power spectrum well up to $k\sim 0.2\ihMpc$. Again, despite the larger biases of the MASS cases, they preserve baryonic features no worse than the NUM cases. Figure \ref{fig:dlnPdlnkminusz0.3red} shows the derivatives in redshift space compared to the matter power spectrum in real space. All biased redshift-space power spectra trace the features in the real space matter power spectrum fairly well up to $k\sim 0.2\ihMpc$ but with more degradation in the NUM cases. 

To summarize, subtracting the smooth anomalous power helps to recover the contrast of the baryonic features at $z=0.3$ as well. With small biases ($b \lesssim 2$), the recovered contrast is comparable to the contrast in underlying matter power spectra even in quasilinear scales meaning that the nonlinear scales deduced in \S~\ref{subsec:nonlinminus} is valid despite biasing.  Also the recovered contrasts do not seem very sensitive to moderate variations of biases, which is consistent with the results at $z=1$.

\subsection{Bias effects at $z=3$} 
\begin{figure*}[t]
\epsscale{2}
\plottwo{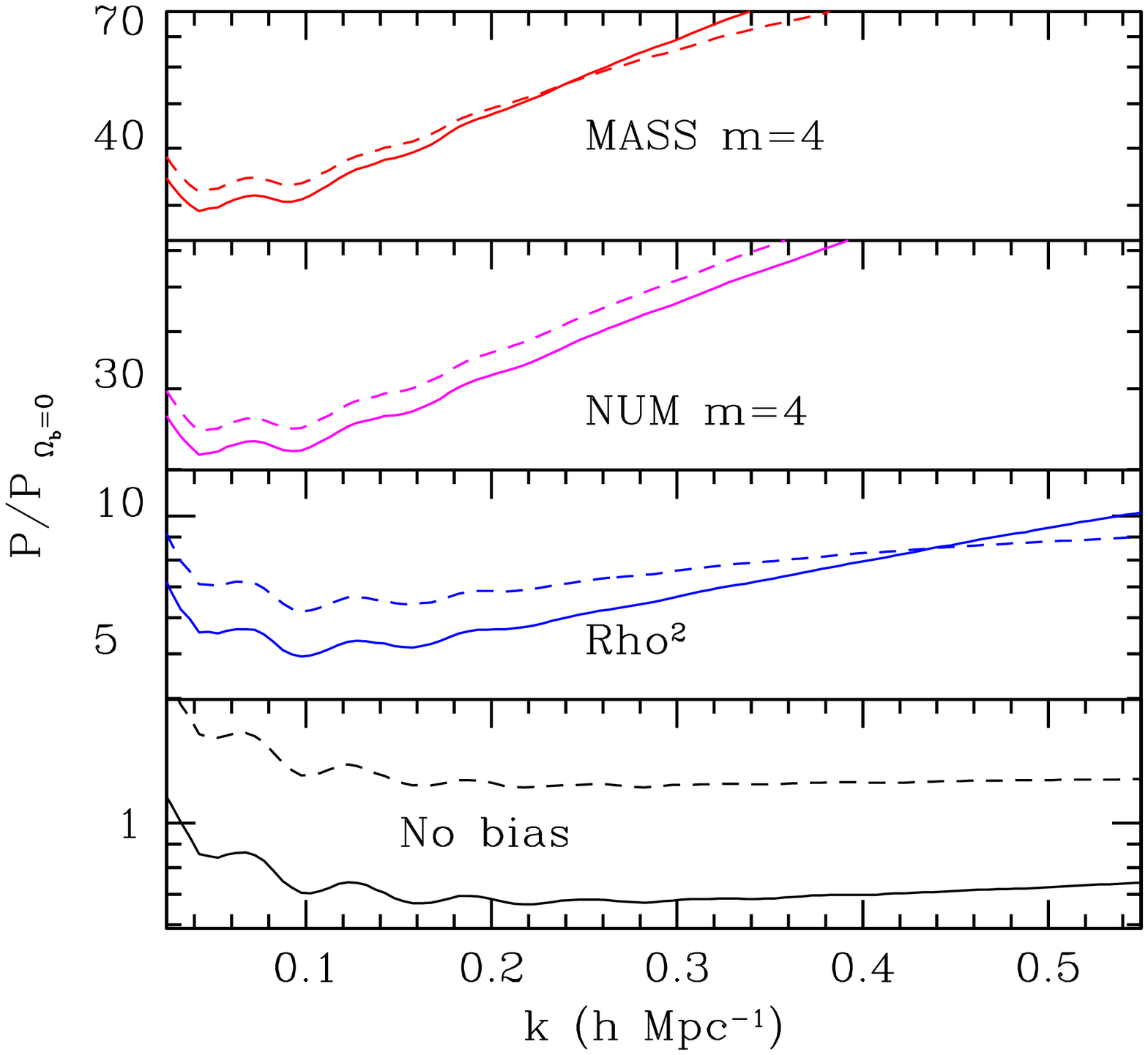}{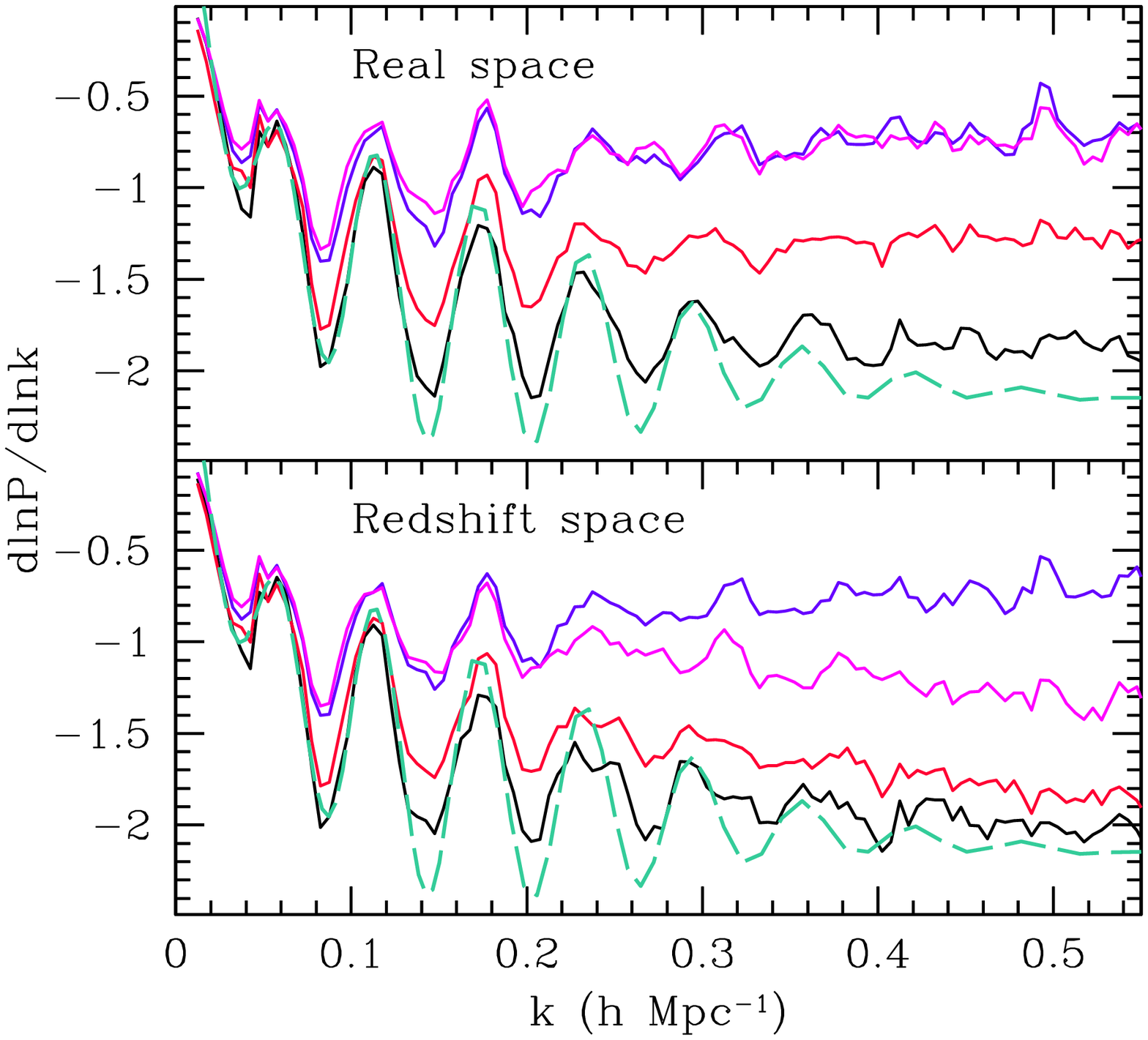}
\caption{Biased power spectra at $z=3$. Left: biased power divided by a zero-baryon power spectrum. Solid lines are for the real space clustering, and dashed lines with the same color are for the corresponding redshift-space clustering. Right: $d\ln P/d\ln k$ from the biased power spectra. Green dashed line: the input power spectrum, black: the matter power spectrum at $z=3$, red: $\Ro$, blue: NUM with $m=4$ and violet: MASS with $m=4$.  }
\label{fig:T0bz3}
\epsscale{1}
\end{figure*}

We next show the results at our highest redshift bin, $z=3$. The number of simulation boxes used for this redshift is 30, which is smaller than the other redshift bins. We generated three biased tracers: $m=4$ for MASS, $m=4$ for NUM, and $\Ro$ (Figure \ref{fig:T0bz3}).  The former two cases generate power spectra with $b \sim 5.5$ and 4.9, which are too high for Lyman break galaxies \citep{Steidel96}. Correspondingly, the number density of these halos is very small, leading to significant noise in the power spectra. The $\Ro$ model, on the other hand, generates a bias of $b \sim 2.5$, similar to that of Lyman break galaxies (LBG). The anomalous power of $\Ro$ is $70\%$ of the shot noise effect from the number density of $10^{-3}h^3 {\rm\;Mpc}^{-3}$ that we assumed for the sample in  SE03 although the relative effect is nearly equivalent. The relative anomalous power amplifies the power by a factor of two at $k \sim \kmax (=0.53\ihMpc)$ for the $\Ro$ case.  

\begin{figure*}[t]
\epsscale{2} 
\plottwo{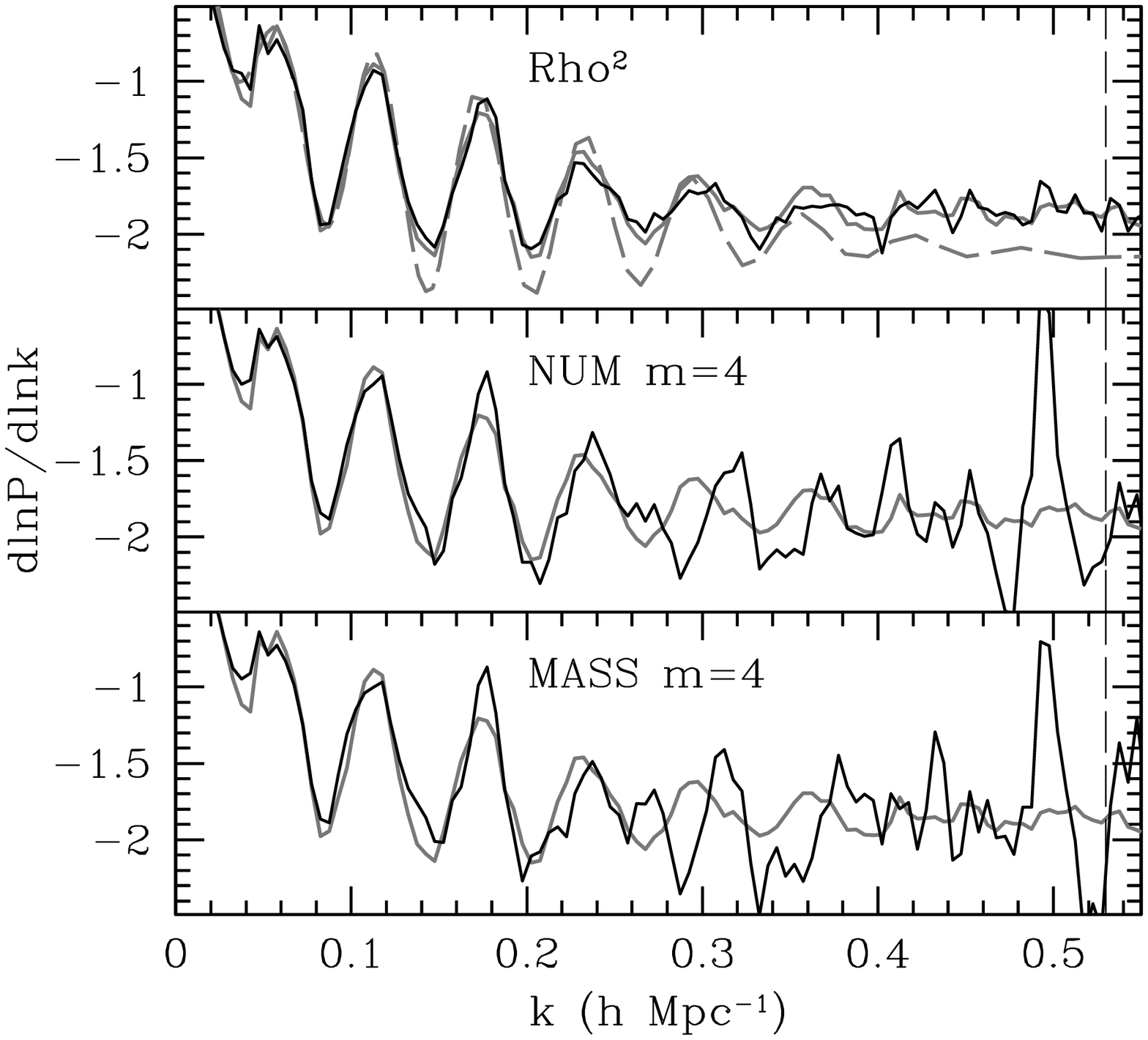}{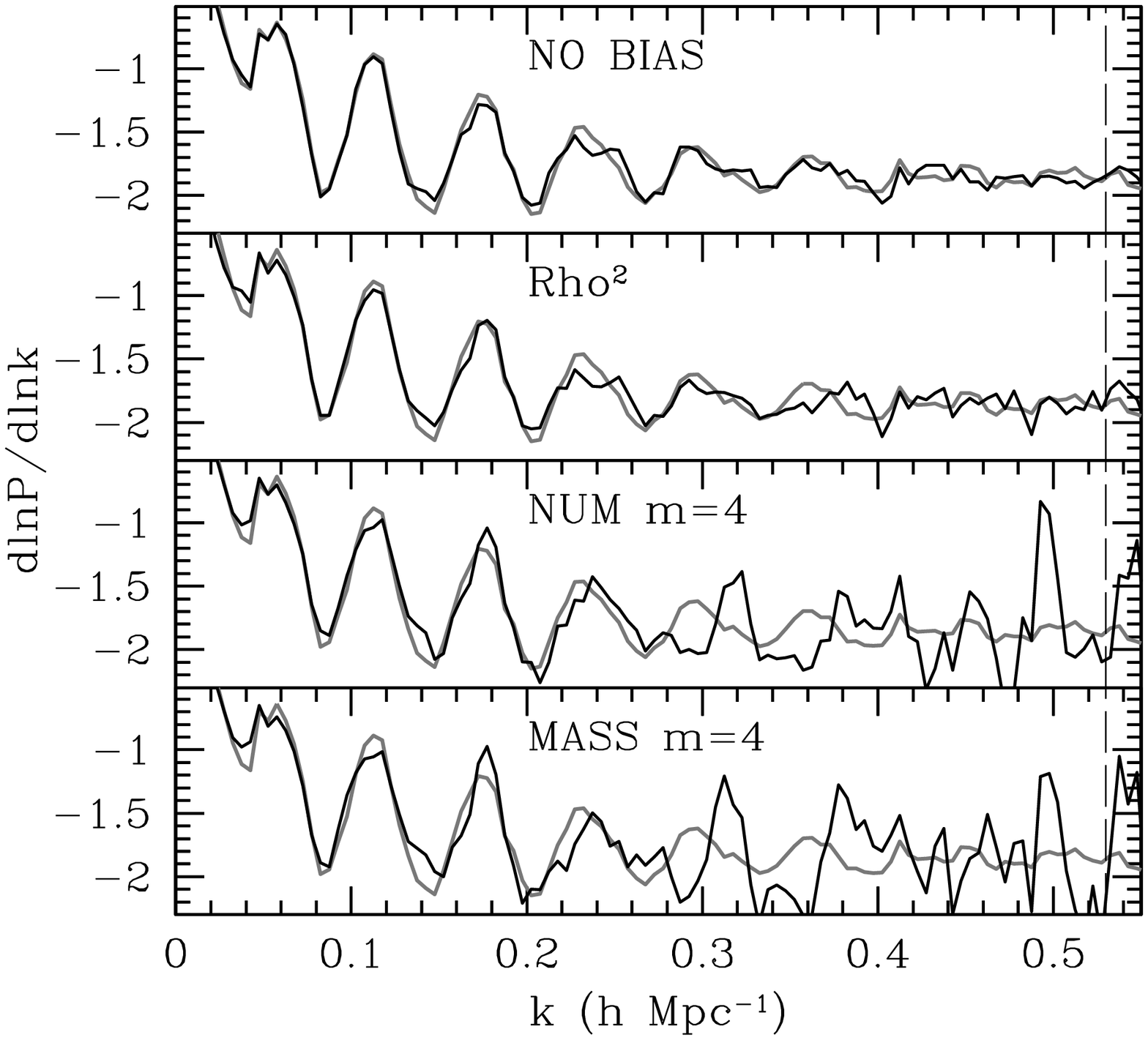}
\caption{$d\ln P/d\ln k$ from $P_{\rm biased}-\fNL(c_0,k,k^2)$ at $z=3$ in real space and redshift space. The $\Ro$ case and the MASS case in redshift space are corrected for the finger-of-God suppression beforehand. Left: the real-space power spectra. Gray dashed line: the input power spectrum in real space, gray solid line: the matter power spectrum in real space, black solid lines: the biased power spectra in real space. Right: the redshift-space power spectra. Gray solid line: the matter power spectrum in {\it real} space, black solid lines : the biased power spectra in {\it redshift} space. Note that the y-axes in the left and the right panels are not scaled the same unlike the previous figures. The vertical line is at $\kmax$ of $0.53 \ihMpc$. From the figure, the tracers with a very large bias do not mimic the underlying matter power spectra very well. One sees that the mildly biased $\Ro$ case ($b \sim 2.5$) probes baryonic features up to $k \sim 0.35\ihMpc$. }
\label{fig:dlnPdlnkminusz3}
\epsscale{1}
\end{figure*}

While the matter power spectrum preserves baryonic peaks up to $k\sim 0.4\ihMpc$ (Figure \ref{fig:dlnPlin}), Figure \ref{fig:dlnPdlnkminusz3} shows that even the $\Ro$ case cannot probe baryonic features beyond $k \sim 0.35\ihMpc$ either in real space (left panel) or redshift space (right panel). The MASS and NUM cases trace the acoustic oscillations only up to $k \sim 0.25\ihMpc$. That is, the tracers with very large bias do not mimic the underlying matter power spectra very well, unlike tracers with moderate biases at lower redshift bins. This is likely because of the statistical noise from a small number of high mass halos, but we cannot exclude the possibility of an emergence of mode coupling as the nonlinear bias effect becomes very large.

We estimate that galaxies with $b \sim 3$ at this redshift will recover baryonic features up to wavenumber of about $0.3 \ihMpc$. The details of the result may vary depending on the biasing schemes and the shot noise.

\subsection{Summary of the effect of bias}

To summarize the general effect of bias on baryonic features, subtracting a smooth function to match the slope of the matter power spectrum largely restores the baryonic features in the underlying matter power spectra if the amplitude of bias is moderate. This implies that a moderate anomalous power of bias whether from shot noise or a nonlinear bias effect does not erase the initial features. In other words, it is reasonable to assume that the anomalous power is smooth in wavenumber and does not generate features that mimic baryonic oscillations.

We find that the detailed effects of bias not only scale with the amplitude of bias and the anomalous power but also depend on the biasing schemes used. For very large amplitudes of bias, we clearly lose information whether it is due to increased nonlinear bias effects or merely shot noise.

Based on the baryonic features preserved in the matter power spectra (\S~\ref{sec:matterps}) and the effect of bias, we summarize the nonlinear scales in the biased power spectra. At $z=0.3$, the results are encouraging in that moderately biased power spectra ($b \lesssim 2$) preserve baryonic features at $k< \kmax$ ($0.11 \ihMpc$) and record a fair amount of linear information even beyond $\kmax$ in both real and redshift space. The biased power spectra at $z=1$ ($b \lesssim 2$) show decreased contrast for $k< \kmax$ ($0.19 \ihMpc$) and contain attenuated traces of baryonic features beyond $\kmax$. At $z=3$, the biased power spectrum with $b\sim 3$ will preserve the features up to $k \sim 0.3 \ihMpc$, but no further. 

The biased power in redshift space traces the real-space biased power reasonably well with partial degradations in the baryonic features depending on biasing schemes. Suppressing the finger-of-God effect does not help to preserve real-space features any better, implying that the motions between halos do not strictly respect linear theory.  

To this point, we have shown the successful restoration of baryonic features in linear and quasilinear scales from various biasing schemes based on halo-mass thresholds. Given that all mass thresholds preserve the features, it is unlikely that more complex descriptions of galaxy populations, e.g., superpositions of our biasing schemes, would change the results. Similarly, a more stochastic local bias would not likely remove the oscillations, although stochastic models can increase anomalous power \citep{Dekel99,Sch98} thereby reducing contrast and increasing noise.

\section{Impacts on cosmological distance estimation : $\chi^2$ analysis of the N-body data}\label{sec:chisq}

We next consider the impact of nonlinearity and bias on the statistical constraints on cosmological distances from the baryonic features. We are interested in using the power spectrum measurement to constrain a distance scale, and therefore we want to estimate how well we can constrain dilations in wavenumbers. We define this dilation parameter as $\alpha(= k_{\rm ref}/k_{\rm true})$. The error on the dilation parameter $\alpha$ represents the errors on the angular diameter distance $\DA$ and Hubble parameter $\hz$. We perform a $\chi^2$ analysis to fit the spherically averaged power spectra in real space from N-body simulations, $P_{\rm obs}$, to a linear combination of the input linear power spectrum, $P_{\rm linear}$, and an additional polynomial function (eq.\ [\ref{eq:chi}]). The fit parameters are $\alpha$, a multiplicative bias $b_0$, a scale-dependent bias $b_1$, and additive terms for nonlinear growth or an additional constant ($a_0$, $a_1$ and $a_2$). The mean value of $\alpha$ is expected to be unity since we set the reference cosmology to be equal to the true cosmology for simplicity. 

\begin{equation}\label{eq:chi}
P_{\rm obs}(k_{\rm ref})=(b_0+b_1 k_{\rm ref})\times P_{\rm linear}(k_{\rm ref} /\alpha)+(a_0+a_1 k_{\rm ref}+a_2 k_{\rm ref}^2) 
\end{equation}

We try two cases, with and without including $b_1$. In both cases, we fit power at wavenumbers beyond $\kmax$ to set the nonlinear trends. We note that even though the model spectrum contains higher harmonics of the baryon oscillations that may be absent from the data, this need not bias the distance measurement. The smooth portion of the nonlinear power spectrum that replaces higher harmonics does not have narrowband features to match on the linear model and hence impose a preferred physical scale. We will return to this point later in this section.

The mean value and error of $\alpha$ are computed using jack-knife subsampling of the simulations \citep{Lupton93}. Since we do not know the true covariance matrix, we assume a Gaussian error in each $k$-bin constructed from the power spectrum averaged over all sets. 
Although assuming a Gaussian error implies that we underestimate the statistical noise relative to the true non-Gaussian error, the variations among jack-knife estimates of $\alpha$ in the 51 subsamples (30 at $z=3$) should reflect the non-Gaussian, mode-coupled error as these subsamples are drawn from actual nonlinear N-body data. In other words, our fitting slightly misweights the data relative to optimum but should not produce overly optimistic $\sigma_{\alpha}$ compared to the true error.

The behavior of resulting errors are consistent with the effect of nonlinearity and bias that we studied in the previous sections. 
For the underlying matter power spectra at $z=3$, we derive $\sigma_{\alpha}\sim 0.35 (0.36)\%$ with wavenumbers less than $\kfit=0.3 (0.5) \ihMpc$. For the $\Ro$ biased power spectrum, $\sigma_{\alpha}\sim 0.35\%$ when $\kfit \sim 0.3 \ihMpc$. Increasing the range of $k$ beyond this value increases the error, which is consistent with the noisy feature in the left panel of Figure \ref{fig:dlnPdlnkminusz3}. If we scale the error to the survey volume assumed in  SE03, this error value corresponds to $1\%$. The anomalous power in the $\Ro$ case is close to the shot noise we assumed there. The analytic results in  SE03 implies $\sigma_\alpha \sim 0.93\%$ (from $1/\sigma^2_{\alpha}=1/\sigma^2_{\DA}+1/\sigma^2_{\hz}$), in good agreement. 

At $z=1$, we calculate $\sigma_\alpha \sim 0.4\%$ for the underlying nonlinear matter power spectrum if we include a region up to $\kfit = 0.3\ihMpc$. Among the biased power spectrum, the NUM case with $m=4$ has a bias value close to that of the redshift bin at $z=1$ in SE03. For this case, $\sigma_\alpha \sim 0.4-0.5\%$, which corresponds to $\sigma_\alpha \sim 1.6-1.8\%$ when scaled to the baseline survey volume in  SE03. 
This is to be compared with $\sigma_\alpha$ of $1.4\%$ for the corresponding number density in SE03. 
Thus the simulation indicates a slightly worse precision relative to our previous predictions in SE03. The equivalent $\kmax$ for a linear approximation will be $0.17-0.18 \ihMpc$ at $z=1$, rather than $0.19 \ihMpc$.

At $z=0.3$, we find more optimistic results, as we would expect from Figure \ref{fig:dlnPdlnkminusz0.3}. For the underlying nonlinear matter power spectrum, we get $\sigma_\alpha \sim 0.6\%$ for $\kfit=  0.3\ihMpc$. For the MASS cases with $m=10$ and $m=30$, which are similar to LRG samples, we find $\sigma_\alpha \sim 0.8-0.9\%$. When scaled for a survey volume of $1 \trihGpc$, $\sigma_\alpha$ is $2.1-2.3\%$ (the equivalent $\kmax \sim 0.15-0.155 \ihMpc$), which is better than the $3.9\%$ calculated from values in SE03. 
This is to be compared with the current observations: the $4\%$ measurements from \citet{Eis05} would give $\sigma_\alpha \sim 3\%$ when scaled to $1 \trihGpc$. The cause of the difference between $2.1-2.3\%$ and $3\%$ is due to the neglect of redshift distortions in this modeling.

In general, the errors calculated with and without $b_1$ are consistent. Without $b_1$, the mean values of $\alpha$ are close to unity, indicating negligible bias. But $\alpha$ is slightly biased above 1 in cases with $b_1$, particularly at lower redshift, albeit by $< 1\%$ in most of cases. Since we use the linear power spectrum to match nonlinear power, the fitting process favors a negative $b_1$ to match the erased baryonic features and, in order to compensate the resulting phase shifts of the oscillations, biases $\alpha$ preferentially above $1$. On the other hand, without $b_1$, the fitting process does not have means to use bias on $\alpha$ to account for the erased features, giving little bias on $\alpha$. Including an appropriate recipe to account for the erasure of the baryonic features should remove this bias. That is, this bias would likely be easy to calibrate and remove using N-body simulations of a reasonable cosmological model. 

The results of $\chi^2$ analysis can be translated to the survey volumes required to achieve $\sigma_\alpha \sim 1 \%$. Table \ref{tab:equivol} presents estimates of the required survey volumes, assuming biased power spectra in real space.

We do not extend the $\chi^2$ analysis to the redshift-space power spectra because of our lack of a reliable model of the redshift distortions to fit. There are deviations from the Kaiser formula on large scales as well as the finger-of-God effect on intermediate and small scales, both of which could involve an arbitrary angular dependence in two dimensions (reduced from three dimensions by the azimuthal symmetry). 
Nevertheless, the comparisons between the real-space and redshift-space power spectra in our study lead to qualitative estimations of the effects of redshift distortions. Due to the decreased contrast of the baryonic features we observed in the spherically averaged power spectra in redshift space, we expect that the errors on $\hz$ will be degraded relative to the analytic prediction. At $z=3$, we expect the effect will be insignificant. At $z=1$, the degradation will produce a larger error on $\hz$ than in SE03. At $z=0.3$, a degradation due to the redshift distortions will increase $\sigma_H$ but likely no worse than the estimates in SE03. 
It is important to note that the information in the spherically averaged redshift-space power spectra will not be quantitatively equivalent to that in the two-dimensional redshift-space power spectra. Here we are averaging out the nonlinear redshift distortions, which are actually angle-dependent.
For example, a simple exercise of $\chi^2$ analysis using only the modes nearly along the line of sight suggests that the degradation of $\hz$ at $z=1$ could be as large as a factor of two between real space and redshift space. We plan to investigate this in a future study.  

\section{Conclusion}

We have used a large set of N-body simulations to show that the baryonic oscillations from the large galaxy redshift surveys survive on large scales well despite the mild nonlinearity of gravity, redshift distortions, and bias. We compared the nonlinear effect on the baryonic features observed in the N-body results with the choices of nonlinear scale $\kmax$ in SE03.

As expected, the nonlinear gravitational evolution erased the baryonic features progressively from smaller scales to larger scales as the redshift decreased. In real space, the nonlinear scales we have assumed in  SE03 seem fairly conservative at $z=0.3$ and $z=1$. We find that we need a slightly smaller $\kmax$ for $z=3$, but this modification has a minor effect on standard ruler test due to the small nonlinear scale. The redshift distortions imposed an additional obscuration for $k< \kmax$. Nevertheless, the redshift-space power spectra reasonably traced the features in the real-space power.

We have shown that moderate nonlinear bias ($b < 3$) does not erase the initial features. The biased power spectra follow the features in the underlying matter power spectrum fairly well once the broadband shape is restored. The effect of bias is not only proportional to the amplitude of the bias and anomalous power but also depends on the biasing models as well. However, these dependences seem to have minor effect on the underlying baryonic features. In redshift space, suppressing the finger-of-God effect does not improve recovery of the features, and this indicates that nonlinear effects in the velocity fields on large scales obscure baryonic features as well. Nevertheless, the redshift-space biased power spectrum reproduces the baryonic features of the real-space counterpart reasonably well despite the degradation mainly due to the modes along the line of sight.

From $\chi^2$ analysis of N-body results in real space, we predict errors on cosmological distances similar to those in SE03. Thus the nonlinear scale $\kmax$ we have adopted in SE03, with minor modification detailed below, adequately describes the effect of nonlinearity on the standard ruler test.  Furthermore, this implies that the cosmological distortions will be indeed distinguishable from nonlinear growth and scale-dependent bias, and so the derived uncertainty on cosmological distances depends on the degree of erasure of the baryonic features.
Considering nonlinear gravity, redshift distortions and the clustering bias effect from N-body results all together, we estimate nonlinear scales appropriate for calculation of the information in the baryonic features. We consider both the loss of information for $k< \kmax$ and the additional linear information on smaller scales which compensates the loss. Referring to the results from the $\chi^2$ analysis, the biased power spectra at $z=3$ with $b\sim 3$ traces baryonic features well up to $k\sim 0.3\ihMpc$ both in real and redshift space. At $z=1$ with $b\lesssim 2$, $\kmax \sim 0.17-0.18\ihMpc$ are appropriate in real space, but we need a slightly smaller $\kmax$ for redshift space. At $z=0.3$ with $b \lesssim 2 $, $\kmax \sim 0.15\ihMpc$ for real space but smaller $\kmax (> 0.11\ihMpc)$ if we consider redshift distortions. We translate these results to the distance measurements: at $z=3$ and $z=0.3$, we expect that $\DA$ will be constrained as well as the estimates in SE03 while $\hz$ will be slightly less well constrained because of the nonlinear redshift distortion effect on the baryonic features. At $z=1$, the deviation will be the largest. While $\DA$ will be near the estimates in SE03, $\hz$ can be as large as twice of $\DA$. 

To summarize, the standard ruler test using baryonic features are robust against nonlinear effects in the linear and quasilinear regime. Therefore, using the standard ruler, the on-going and future large galaxy redshift surveys will measure the dilations in observed scales due to cosmological distortions at various redshifts to excellent accuracy, providing a superb probe of the acceleration history of the universe. 

\acknowledgements
We thank Martin White for helpful discussions.  We were supported by grant AST-0098577 and AST-0407200 from the National Science Foundation. DJE is further supported by an Alfred P.\ Sloan Research Fellowship.

\clearpage
\onecolumn

\begin{table}
\begin{center}
\tablewidth{0pt}
\tabletypesize{\small}
\caption{\label{tab:equivol}Required Survey Volume to Achieve $\sigma_\alpha = 1 \%$}
\begin{tabular}{l|ccc}
\tableline\tableline
Redshift&bias& $\Vsur (\trihGpc)$& $n_{\rm eff} (h^3 {\rm\;Mpc}^{-3})$\\ \hline
0.3&2.0&4.4&$2.1 \times 10^{-4}$\\
&2.3&5.5&$1.2 \times 10^{-4}$\\
1&1.7&1.4 &$2.1  \times 10^{-3}$\\
&2.4&1.7 &$3.4  \times 10^{-4}$ \\
3&2.5&0.5&$1.4 \times 10^{-3}$ \\
\tableline
\end{tabular}
\tablenotetext{}{ Note. -- The approximate survey volumes required to achieve $\sigma_\alpha = 1 \%$ for the biased power spectra. Note that the values are based on our $\chi^2$ analysis of the real space power spectra, and this means that redshift distortions are not accounted for. The effective number density $n_{\rm eff}$ is the inverse of the anomalous power at $k \sim 0$.}
\end{center}
\end{table}

\end{document}